\documentclass[12pt]{article}
\usepackage{mathtools,amssymb,amsfonts,geometry,xcolor,authblk, float, multirow, tikz, dynkin-diagrams, cite}
\numberwithin{equation}{section}
\usepackage[linktoc=all]{hyperref}
\hypersetup{
	colorlinks,
	citecolor=black,
	filecolor=black,
	linkcolor=black,
	urlcolor=black
}
\usepackage[font=small,labelfont=bf]{caption}
\usepackage[font=small,labelfont=bf]{subcaption}

\usetikzlibrary{shapes, positioning, chains, arrows, fit, decorations.pathreplacing, shapes.misc, decorations.markings }
\tikzstyle{every picture}+=[remember picture]
\tikzstyle{na} = [baseline=-.5ex]
\def\ns#1{
	\node[circle, draw, fill=white] at (#1){};
	\node[cross out, draw] at (#1){};
}

\def\ns#1{
	\node[circle, draw, fill=white,minimum size=0.6cm] at (#1){};
	\node[cross out, draw,minimum size=0.425cm] at (#1){};
}

\newcommand{\Ch}{\mathrm{Ch}}
\newcommand{\PE}{\mathrm{PE}}
\newcommand{\diff}{\mathrm{d}}

\newcommand{\Ncal}{\mathcal{N}}

\newcommand{\R}{\mathbb{R}}
\newcommand{\T}{\mathbb{T}}

\newcommand{\surm}{\mathrm{SU}}
\newcommand{\urm}{\mathrm{U}}

\newcommand{\orm}{\mathrm{O}}
\newcommand{\sorm}{\mathrm{SO}}
\newcommand{\sprm}{\mathrm{Sp}}
\newcommand{\spn}{\mathrm{Sp}(N)}
\newcommand{\sh}{\mathrm{sh}\,}
\newcommand{\ch}{\mathrm{ch}\,}

\title{D-type Minimal Conformal Matter: \\Quantum Curves, Elliptic Garnier Systems, and the 5d Descendants}
\author[1]{Jin Chen\,}
\author[2]{Yongchao L\"u\,}
\author[3]{Xin Wang\,}
\affil[1]{\small\it  
	 Department of Physics, Xiamen University, Xiamen, 361005, China}
 \affil[ ]{{\it e-mail}: zenofox@gmail.com}
\affil[2]{\small\it  
	School of Physics, Korea Institute for Advanced Study, Hoegiro 85, Seoul 02455, Korea}
\affil[ ]{{\it e-mail}: lychaoaa@gmail.com}
\affil[3]{\small\it
	Quantum Universe Center, Korea Institute for Advanced Study, Hoegiro 85, Seoul 02455, Korea}
\affil[ ]{{\it e-mail}: wxin@kias.re.kr}
\date{}

\begin{document}
\noindent
\hspace{\fill} KIAS-Q23004\\

\begingroup
\let\newpage\relax
\maketitle
\endgroup

\begin{abstract}	
We study the quantization of the 6d Seiberg-Witten curve for D-type minimal conformal matter theories compactified on a two-torus. The quantized 6d curve turns out to be a difference equation established via introducing codimension two and four surface defects. We show that, in the Nekrasov-Shatashvili limit, the 6d partition function with insertions of codimension two and four defects serve as the eigenfunction and eigenvalues of the difference equation, respectively. We further identify the quantum curve of D-type minimal conformal matters with an elliptic Garnier system recently studied in the integrability community. At last, as a concrete consequence of our elliptic quantum curve, we study its RG flows to obtain various quantum curves of 5d $\sprm(N)+N_f \mathsf{F},N_f\leq 2N+5$ theories. 
\end{abstract}

\newpage

\tableofcontents


\section{Introduction}
\label{sec:intro}
Supersymmetric gauge theories with eight supercharges have played a prominent role in understanding non-perturbative physics in the strongly coupled regime. One of the most remarkable achievements in this area can be traced back to the 1990s when Seiberg and Witten studied the quantum vacuum structures of 4d $\mathcal N=2$ gauge theories \cite{Seiberg:1994rs,Seiberg:1994aj}. They discovered that the quantum vacua are solely captured by an algebraic curve, and the low-energy effective theory is described by a holomorphic function. The curve and the holomorphic function are now known as the Seiberg-Witten curve (SW-curve) and prepotential, respectively. The low-energy BPS spectra, with both perturbative and instanton contributions, of the theories in the far infrared, are encoded in the periods of the SW-curves. Later on, another breakthrough was provided in \cite{Nekrasov:2002qd} where the Seiberg-Witten prepotential can be presented by a two-parameter generalization known as Nekrasov partition function, from which the corresponding SW-curve can be re-derived in a thermal dynamical limit and proven in \cite{Nekrasov:2003rj,Nakajima:2003pg,Nakajima:2009qjc}. In this framework, the curve is also interpreted as the phase space of a codimension-two surface defect of the theory. Along this line, it has been realized that the SW-curve can be quantized by setting one of the deformation parameters to zero in the Nekrasov partition function \cite{Nekrasov:2009rc}. The quantum curve thus turns out to be an operator that annihilates the expectation value of the codimension-two defect operator. The above 4d story can be further extended to 5d and 6d superconformal field theories (SCFTs) living on $\mathbb R^4\times \mathbb S^1$ and $\mathbb R^4\times \mathbb T^2$ respectively. The algebraic curves are thereby lifted to hyperbolic and elliptic ones. The \emph{elliptic quantum curves} are particularly interesting to study to understand the moduli space of various 6d SCFTs compactified on a torus, and thus shed new light on their properties.

On the other hand, it has been shown that nontrivial interacting superconformal field theories (SCFTs) can exist only in spacetimes with a maximum of six dimensions \cite{Nahm:1977tg}. As a result, 6d SCFTs can be regarded as the mother theories of all supersymmetric quantum field theories (QFTs) in lower dimensions that arise from compactification on various manifolds. For instance, the classification of 5d theories has been explored by compactifying them on a circle \cite{Bhardwaj:2018yhy,Bhardwaj:2018vuu,Apruzzi:2018nre,Apruzzi:2019vpe,Apruzzi:2019opn,Bhardwaj:2019fzv,Bhardwaj:2020gyu,Bhardwaj:2020kim,Bhardwaj:2019xeg, Braun:2021lzt}. In recent years, significant progress has been made in classifying 6d $\mathcal N=(1,0)$ SCFTs from F-theory compactified on elliptic fibered Calabi-Yau threefolds \cite{Heckman:2013pva,Heckman:2015bfa,DelZotto:2017pti}. From this vast landscape of 6d SCFTs, one can further compactify them to lower dimensional supersymmetric QFTs and investigate many intriguing non-perturbative properties therein. A particularly interesting example is the compactifications of 6d $\mathcal N=(2,0)$ SCFTs on Riemann surfaces punctured by codimension two defects, resulting in a wide range of 4d $\mathcal N=2$ theories known as class $\mathcal S$ theories and their dualities \cite{Gaiotto:2009we,Gaiotto:2009hg,Alday:2009fs}. More recently, the construction is generalized to 6d $\mathcal N=(1,0)$ down to 4d $\mathcal N=1$\cite{Gaiotto:2015usa,Razamat:2016dpl,Ohmori:2015pua,Ohmori:2015pia,Kim:2017toz,Bah:2017gph,Kim:2018bpg,Razamat:2018gro,Kim:2018lfo,Ohmori:2018ona,Razamat:2019mdt,Chen:2019njf,Pasquetti:2019hxf,Razamat:2019ukg,Razamat:2020bix,Hwang:2021xyw,Nazzal:2021tiu, Razamat:2022gpm, Heckman:2022suy, Kim:2023qbx}, and novel connections have been established between elliptic quantum difference equations and surface defects introduced in the 4d theories in terms of their superconformal index \cite{Gaiotto:2012xa,Gaiotto:2015usa,Nazzal:2018brc,Nazzal:2021tiu,Nazzal:2023bzu}. Meanwhile, in the context of 6d $\mathcal N=(1,0)$ or their corresponding KK theories, it has been realized that, in the case of rank one 6d theories on the tensor branch with trivial gauge groups, the elliptic quantum difference equations studied in the 4d setup are precisely the elliptic quantum SW-curves of the corresponding 6d theories, and the surface defects in 4d also become a class of important codimension four observables, the Wilson surface defects, that serve the eigenvalues of the quantum curves\cite{Bullimore:2014awa, Chen:2020jla, Chen:2021ivd}. Furthermore, the elliptic quantum curves remarkably bridge another interesting field of mathematical physics, the elliptic integrable systems, where the quantum curves are identified there as the spectral curves of the associated integrable systems. Therefore, it naturally motivates us to develop a systematic approach to study 6d SW-curves from the supersymmetric defects and their partition functions, and also understand their intriguing connections to the elliptic quantum systems.

In this paper, we will continue our exploration on 6d Seiberg-Witten curves by investigating the codimension two and four supersymmetric defects and their partition functions on the tensor branch. We focus on the 6d D-type minimal conformal matters (CM), which describe the low-energy dynamics of a single M5-brane probing $D_{N+4}$-type singularity. These theories are a direct generalization of the E-string theory, as studied in \cite{Chen:2021ivd}. It also can be realized as a $\sprm(N)$ gauge theory with $2N+8$ fundamental flavors on a -1 curve in the framework of F-theory, where the E-string can be regarded as a ``$\sprm(0)$'' theory. Therefore one can Higgs a $\sprm(N)$ theory all the way downwards to the E-string, as well as inserting half-BPS codimension two and four defects in the same fashion of \cite{Chen:2020jla, Chen:2021ivd, Chen:2021rek}. Using the techniques of localization, one can study the vacua moduli of the 6d theories and compute their instanton string partition functions in presence of various defects in the Nekrasov's $\Omega_{\epsilon_{1,2}}$-background. In a further Nekrasov-Shatashvili limit $\epsilon_2\rightarrow 0$, one can establish the quantum SW-curve, for the D-type minimal CM, that acts on the codimension two defect partition function served as a wave function $\Psi_{\rm inst}(x;\epsilon_1)$ and generate the codimension four defect partition function $\chi_{\rm inst}(x;\epsilon_1)$ as the eigenvalue of the curve:
\begin{align}
\mathcal D_{{\rm inst}}\, \Psi_{\rm inst}(x;\epsilon_1)=\chi_{\rm inst}(x;\epsilon_1)\, \Psi_{\rm inst}(x;\epsilon_1)\,,
\end{align}
with
\begin{align}
\mathcal D_{{\rm inst}}\equiv Y+\frac{\mathfrak q^2 }{\vartheta_1(2x)\vartheta_1(2x+\epsilon_1)^2\vartheta_1(2x+2\epsilon_1)}\frac{\prod_{f=1}^{2N+8}\vartheta_1\left(x\pm m_f+\frac{\epsilon_1}{2}\right)}{\prod_{i=1}^N\vartheta_1(x\pm\alpha_i)\vartheta_1(x\pm\alpha_i+\epsilon_1)}\cdot Y^{-1}\,,\notag
\end{align}
where $Y$ is the difference operator satisfying $Y\cdot X=Y\cdot e^x=e^{-\epsilon_1}X\cdot Y$. The quantum curve, or say the difference operator, can be Higgs down to the E-string one. In the classical limit $\epsilon_1\rightarrow 0$, it also returns back to the classical SW-curve studied in \cite{Haghighat:2018dwe}. Therefore $\mathcal D_{\rm inst}$ is proposed as a quantization of the classical 6d Seiberg-Witten curve for D-type minimal conformal matters.

In the case of E-string theory, it has been found that its quantum curve can be remarkably identified as the Hamiltonian of the van Diejen integrable system. Specifically, it has been shown that the 1-instanton contribution of the Wilson surface defect corresponds to the 4-theta external potential in the van Diejen operator. Consequently, one would also expect a relation between the quantum curve of a generic 6d $\sprm(N)$ theory and certain integrable systems. Indeed, in the paper, we will establish a connection between the $\sprm(N)$ quantum curve and a class of {\it{elliptic Gainier systems}} that has been investigated recently in the integrability community. On the other hand, it is also worth mentioning that the van Diejen operator and its generalizations have been also studied in the analysis of surface defects in various 4d $\mathcal N=1$ theories from the compactifications of 6d $\sprm(N)$ onto Riemann surfaces \cite{Nazzal:2018brc,Nazzal:2021tiu,Nazzal:2023bzu}. It would be very interesting to understand if these difference operators can be introduced in the context of a pure 6d setup, and correspond to what kind of codimension two and four defects.

Another interesting application of the 6d quantum curve is that one can study its different deformations under various limits of its parameters. From the physics perspective, these deformations correspond to triggering RG flows from the 6d SCFTs to a hierarchy of 5d theories. As a result, one obtains a series of quantum curves, as we dubbed ``quantum curve cascades", associated with the flowed 5d theories. In the case at hand, when the 6d $\sprm(N-1)$ theory is compactified on a circle, we arrive at the 5d Kaluza-Klein (KK) theory, which can be effectively described by 5d $\mathcal{N}=1$ $\sprm(N)$ with $(2N+6)$ fundamental flavors. By integrating the masses of flavors, the theory can flow to theories with fewer fundamental flavors. Along this line of flows, we obtain quantum curves for all 5d $\mathcal{N}=1$ $\sprm(N)+N_f\mathsf{F}$ theories with $N_f\leq 2N+5$ in Section \ref{sec:RG}. Especially, by properly tuning these mass parameters, various quantum curves of 5d $\mathcal N=1$ non-Lagrangian theories can be achieved. We use the example of $\mathbb{P}^2\cup\mathbb{F}_6$ \cite{Jefferson:2018irk} to illustrate this point in the subsection \ref{4.6}. In addition, all of the curves serve the quantum version of the classical curves derived from the brane diagrams for 5d $\sprm(N)$ theories in \cite{Hayashi:2017btw,Li:2021rqr}. 

The paper is organized as follows. In Section \ref{2}, we will introduce the 6d $\sprm(N)$ theory and its codimension two and four surface defects. Using its brane setups, we compute the codimension two and four defect partition functions and establish the $\sprm(N)$ quantum curve from them. In Section \ref{Garnier}, we show that the $\sprm(N)$ curve can be identified to an elliptic Gairnier system. In Section \ref{sec:RG}, we derive the 5d quantum curves for $\sprm(N)+N_f\mathsf{F}$ theories, with $N_f\leq (2N+5)$, the explicit expressions can be found in \eqref{eq:curve5}, \eqref{eq:curve4}, \eqref{eq:curve3} and \eqref{eq:curvepi}. Last, Appendix \ref{A} summarizes the definition of theta functions. Appendix \ref{B} and \ref{C} collect the 2-instanton results for partition functions with codimension two and four surface defects in 6d $\sprm(N)$ theory, and Wilson loop expectation values in various 5d $\sprm(N+1)$ theories with different flavors, respectively.

\section{D-type Minimal Conformal Matters}
\label{2}
In this section, we will derive the elliptic difference operator which quantizes the Seiberg-Witten curve of the D-type minimal conformal matters.  In the same fashion as \cite{Chen:2020jla, Chen:2021ivd, Chen:2021rek}, we will first discuss how to introduce codimension two and four defects under $\Omega$-background and compute the partition functions with defects. In the Nekrasov-Shatashivili limit, we will show how the partition functions with the insertion of codimension two and four defects are related via a linear elliptic difference equation where the associated elliptic difference operator gives the quantum Seiberg-Witten curve. The derivation of the quantum Seiberg-witten curve is a straightforward generalization of \cite{Chen:2021ivd}. We verify it from direct instanton computation up to 2-instanton order.
Throughout this paper, we will use $Z_k^{\rm 6d}$ to denote the string partition function without defects, while using $Z_k^{{\rm 6d/4d}}$ ($Z_k^{{\rm 6d/2d}}$) to denote the string partition function with codimension two (codimension four) defects.

\subsection{Set-ups}
\label{sec:setup}
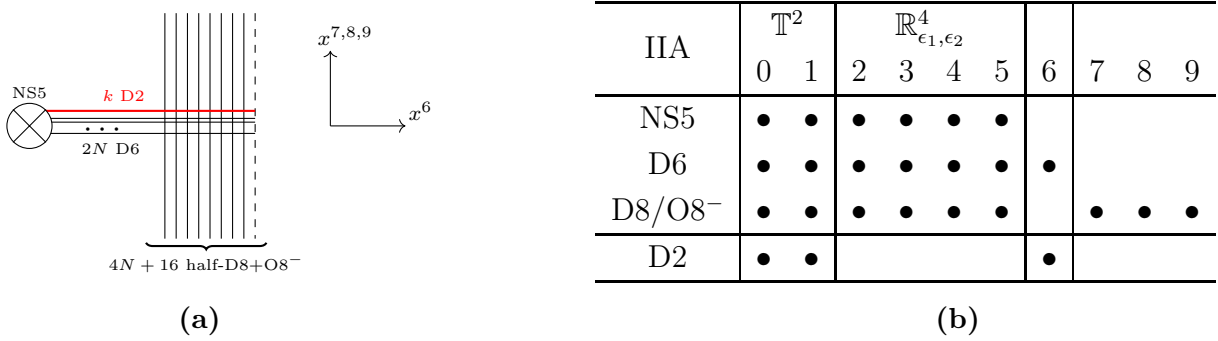
\begin{figure}[H]
\centering
\begin{subfigure}{0.3\textwidth}
\begin{tikzpicture}
     \draw[dashed] (2,1.5)--(2,-1.5);
    %
    \foreach \i in {1,...,8}
    {
    \draw (2-0.15*\i,1.5)--(2-0.15*\i,-1.5);
    }
    \draw (-1,0.1)--(2,0.1);
    \draw (-1,0.05)--(2,0.05);
    \node at (0,-0.05) {$\cdots$};
    \draw (-1,-0.1)--(2,-0.1);
    \draw[red,thick] (-1,0.2)--(2,0.2);
    \ns{-1,0};
    \draw[->] (3,0)--(4,0);
    \draw[->] (3,0)--(3,1);
    \node at (4.2,0.2) {$\scriptstyle{x^6}$};
    \node at (3.2,1.2) {$\scriptstyle{x^{7,8,9}}$};
    \node at (-1,0.45) {\tiny{NS5}};
    \node at (0.08,-0.3) {\tiny{$2N$ D6}};
    \node at (0.25,0.4) {\tiny{\textcolor{red}{$k$ D2}}};
    \draw[decoration={brace,mirror,raise=30pt},decorate,thick](0.55,-0.5) -- node[below=30pt] {\tiny{$4N+16$ half-D8+O$8^-$}  } (2.15,-0.5);
    \end{tikzpicture}
    \caption{}
\end{subfigure}
\hfill
\begin{subfigure}{0.55\textwidth}
\begin{tabular}{c|cc|cccc|c|cccc}
\hline
\multirow{2}{*}{IIA}  & \multicolumn{2}{c|}{$\T^2$} & 
\multicolumn{4}{c|}{$\R^4_{\epsilon_1,\epsilon_2}$} & & \\
  & 0 & 1 & 2 & 3& 4 & 5 & 6  & 7 & 8 & 9 \\ 
\hline
 NS5 & $\bullet$ & $\bullet$  & $\bullet$ & $\bullet$& $\bullet$ & 
$\bullet$&  \\
 D6 & $\bullet$ & $\bullet$ & $\bullet$ & $\bullet$ & $\bullet$ & 
$\bullet$ & $\bullet$
 \\
 D8/O$8^-$ & $\bullet$ & $\bullet$ & $\bullet$ & $\bullet$ & $\bullet$ & 
$\bullet$ & & $\bullet$ & $\bullet$ & $\bullet$
 \\ 
\hline
 D2 & $\bullet$ & $\bullet$ & & & & & $\bullet$ & \\
\hline
\end{tabular}
\caption{}
\end{subfigure}
\caption{Type IIA brane setup comprised of D6-D8-NS5 branes in the presence of an orientifold 8-plane leads to a 6d $\Ncal=(1,0)$ $\sprm(N)$ gauge theory with $2N+8$ fundamental flavors and one tensor multiplet. The case with $N=0$ is the E-string theory. The addition of D2 branes captures the dynamics of the self-dual strings.}
\label{fig:branes}
\end{figure}

The 6d $D$-type minimal conformal matters can be obtained by using a single M5-brane to probe $D_N$ singularity. The resulting theories are $\sprm(N)$ gauge theories with $2N+8$ flavors on their tensor branches. They also admit various brane realizations, e.g. D6/O$6^+$-D8-NS5 studied in \cite{Kim:2015fxa}, or D6-D8/O$8^-$-NS5 branes systems in \cite{Kim:2014dza}. In accordance with the E-string theory case \cite{Chen:2021ivd}, we will use the latter one, see Fig.\,\ref{fig:branes}, in type IIA string theory, to construct the 6d $D$-type minimal conformal matters. It also turns out to be important for the purpose of the ADHM construction in later sections.\\

One can Higgs a pair of hypermultiplets from $\sprm(N+1)$ down to $\sprm(N)$ gauge theory with $2N+8$ flavors. In the brane setup, Higgsing hypermultiplets implies that the D8 brane and its mirror are moved to infinity. When a D8 brane passes through the NS5 branes, emergent D6 branes are attached between the D8 and NS5 branes due to the Hanany-Witten transition. The D6 brane segments on the left and right-hand side of the NS5 brane can rejoin with each other to form a whole single D6 brane. This D6 brane will be also lifted away when the D8 brane and its mirror move to infinity, see below,
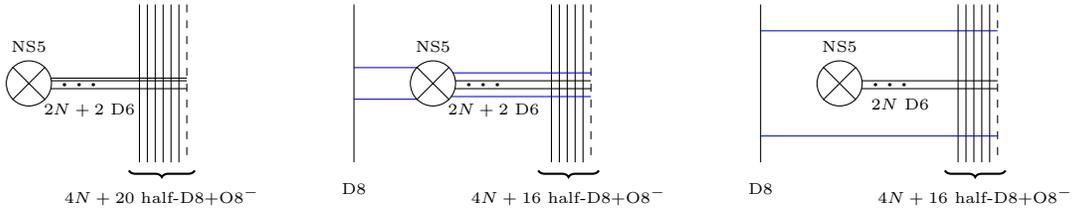
\begin{align}
\begin{gathered}
\begin{tikzpicture}[scale=0.70]
    \draw[dashed] (2,1.5)--(2,-1.5);
    \foreach \i in {1,...,6}
    {
    \draw (2-0.15*\i,1.5)--(2-0.15*\i,-1.5);
    }
    \draw (-1,0.1)--(2,0.1);
    \draw (-1,0.05)--(2,0.05);
    \node at (0,-0.05) {$\cdots$};
    \draw (-1,-0.1)--(2,-0.1);
    \ns{-1,0};
    \node at (-1,0.7) {\tiny{NS5}};
    \node at (0.15,-0.5) {\tiny{$2N+2$ D6}};
    \draw[decoration={brace,mirror,raise=23pt},decorate,thick](0.9,-0.5) -- node[below=25pt] {\tiny{$4N+20$ half-D8+O$8^-$}  } (2.15,-0.5);
    \end{tikzpicture}
\end{gathered}
\quad\quad
\begin{gathered}
\begin{tikzpicture}[scale=0.70]
    \draw[dashed] (2,1.5)--(2,-1.5);
    \foreach \i in {1,...,5}
    {
    \draw (2-0.15*\i,1.5)--(2-0.15*\i,-1.5);
    }
    \draw (-2.5,1.5)--(-2.5,-1.5);
    \draw[blue] (-1,0.3)--(-2.5,0.3);
    \draw[blue] (-1,-0.3)--(-2.5,-0.3);
    \draw[blue] (-1,0.2)--(2,0.2);
    \draw[blue] (-1,-0.25)--(2,-0.25);
    \draw (-1,0.05)--(2,0.05);
    \node at (0,-0.05) {$\cdots$};
    \draw (-1,-0.1)--(2,-0.1);
    \ns{-1,0};
    \node at (-1,0.7) {\tiny{NS5}};
    \node at (0.15,-0.5) {\tiny{$2N+2$ D6}};
    \node at (-2.5, -2) {\tiny D8};
    \draw[decoration={brace,mirror,raise=23pt},decorate,thick](1,-0.5) -- node[below=25pt] {\tiny{$4N+16$ half-D8+O$8^-$}  } (2.15,-0.5);
    \end{tikzpicture}
\end{gathered}
\quad\quad
\begin{gathered}
\begin{tikzpicture}[scale=0.70]
    \draw[dashed] (2,1.5)--(2,-1.5);
    \foreach \i in {1,...,5}
    {
    \draw (2-0.15*\i,1.5)--(2-0.15*\i,-1.5);
    }
    \draw (-2.5,1.5)--(-2.5,-1.5);
    \draw[blue] (-2.5,1)--(2,1);
    \draw[blue] (-2.5,-1)--(2,-1);
    \draw (-1,0.05)--(2,0.05);
    \node at (0,-0.05) {$\cdots$};
    \draw (-1,-0.1)--(2,-0.1);
    \ns{-1,0};
    \node at (-1,0.7) {\tiny{NS5}};
    \node at (0.15,-0.4) {\tiny{$2N$ D6}};
    \draw[decoration={brace,mirror,raise=23pt},decorate,thick](1,-0.5) -- node[below=25pt] {\tiny{$4N+16$ half-D8+O$8^-$}  } (2.15,-0.5);
    \node at (-2.5, -2) {\tiny D8};
    \end{tikzpicture}
\end{gathered}
\,,\notag
\end{align}
where the blue line denotes the to-be-moved D6 branes. 

\paragraph{Perturbative partition function}
For a $\sprm(N)$ gauge theory  with $2N+8$ hypers
\begin{align}
H_\alpha^m=(Y^m_\alpha, (\tilde Y^\alpha_m)^\dagger)\,,\ \ \ \ {\rm with}\ \ \  \alpha=1,\dots, N\,\ \ \ \ {\rm and}\ \ \ m=1,\dots, 2N+8\,,
\end{align}
the $\urm(2N+8)$ global symmetry is enhanced to $\sorm(4N+16)$ due to the pseudo-real property of $\sprm(N)$. Since $\urm(2N+8)\subset \sorm(4N+16)$, one can group the meson operators into three types of representations with respect to $\urm(2N+8)$, i.e.
\begin{align}
\mathcal M^{mn}\equiv J^{\alpha\beta}Y^m_\alpha Y^n_\beta\,,\ \ \ \widetilde{\mathcal M}_{mn}\equiv J_{\alpha\beta}\tilde Y^\alpha_m \tilde Y^\beta_n\,\ \ \ {\rm and}\ \ \ \mathcal N^m_n\equiv Y^m_\alpha \tilde Y^\alpha_n\,,
\end{align}
where $J_{\alpha\beta}$ is the anti-symmetric tensor of $\sprm(N)$. The operators $\mathcal M$, $\widetilde{\mathcal M}$ and $\mathcal N$ correspond to the anti-symmetric, complex conjugate anti-symmetric, and adjoint representations of $\urm(2N+8)$, respectively. Together these three $\urm(2N+8)$ representations form the adjoint representation of $\sorm(4N+16)$. Their fugacities are labeled as
\begin{align}
\mathcal M^{ij} :\, q_1 q_2 e^{-m_i-m_j}\,,\ \ \ \widetilde{\mathcal M}_{ij}  :\, q_1 q_2 e^{m_i+m_j}\,,\ \ \ \mathcal N^m_n  :\, q_1 q_2 e^{-m_i+m_j}\,.
\label{eq:mesons}
\end{align}
Therefore, from the field content of the minimal $D_N$ conformal matters, one can spell out their perturbative contributions to the partition function,
\begin{align}
Z_{\rm pert} \, = \, &Z_{\rm class} \,
\PE\bigg[-\frac{1+ q_1 q_2}{(1-q_1)(1-q_2)(1- p)} \notag \\
& \, \qquad \times \Big(\sum_{i<j}^N\left(A_i A_j+A_i A_j^{-1}+ \left( (A_i A_j)^{-1} +A_i^{-1} A_j \right) p \right)
+ \sum_{i=1}^N\left(A_i^2+A_i^{-2}p \right) \Big) \notag\\
& \,\qquad + \frac{\sqrt{q_1q_2}}{(1-q_1)(1-q_2)(1-p)}\sum_{i=1}^N(A_i+A_i^{-1} p)\sum_{f=1}^{2N+8}\left(M_f+M_f^{-1}\right)\bigg]\,,
\label{eq:perturbative_partition_function}
\end{align}
where $\rm PE$ is the plethystic exponential defined in Appendix \ref{A}, $q_{1,2}=e^{\epsilon_{1,2}}$, $A_i=e^{\alpha_i}$ and $M_f=e^{m_f}$, are the fugacities of $\Omega_{\epsilon_{1,2}}$-background, $\sprm(N)$ gauge multiplet, and hyper multiplets respectively, $p=e^{2\pi i\tau}$ with $\tau$ as the moduli parameter of $\mathbb T^2$, and all other terms independent on $A_i$, e.g. the contribution from tensor multiplet, have been dropped. In eq.\,(\ref{eq:perturbative_partition_function}), we also have applied a flop transition to reverse some of $A_i^{-1}\rightarrow A_i$ that corresponds to assigning $\frac{1}{2}$-BPS boundary conditions when compactifying the 6d theory on $S^1$ to 5d \cite{Chen:2021ivd}. In addition, the classical contribution to the prepotential, denoted by $Z_{\rm class}= \exp\left( \frac{1}{\epsilon_1\epsilon_2}F_{\rm class} \right)$, can be computed from the Green-Schwarz term and one-loop contributions from vector- and hyper-mulplets, which is given by
\begin{align}
F_{\rm class} \, = \, &  \frac{1}{6}\sum_{i<j}^N(\alpha_i\pm\alpha_j)^3+\frac{1}{6}\sum_{i=1}^N(2\alpha_i)^3\notag\\
& \,\, -\frac{1}{12}\sum_{i=1}^{N}\sum_{j=1}^{2N+8}(\alpha_i\pm m_j)^3 +(\phi_0+\frac{\tau}{2})(\sum_{i=1}^{N}\alpha_i^2-\frac{1}{2}\sum_{i=1}^{2N+8}m_i^2)+\cdots.
\end{align}
The ``$\cdots$'' part is irrelevant to our discussion, so we do not list it here. One can check the Higgsing procedure that was diagrammatically discussed before. The perturbative partition function of a $\sprm(N)$ theory can be obtained via Higgsing from a $\sprm(N+1)$ one. More concretely, consider the mesonic operator $\mathcal M^{2N+9,2N+10}$ in eq.\,(\ref{eq:mesons}), with fugacity
\begin{align}
\mathcal M^{2N+9,2N+10}  :\, q_1 q_2 e^{-m_{2N+9}-m_{2N+10}} = \frac{q_1 q_2} {M_{2N+9} M_{2N+10}}.
\label{eq:meson}
\end{align}
The Higgsing is triggered by assigning a non-zero VEV to $\mathcal M$ as
\begin{align}
\left\langle\mathcal M\right\rangle=1\,.
\end{align}
It can be achieved by imposing
\begin{align}
    A_{N+1}=M\,,\quad M_{2N+9}=M\sqrt{ q_1 q_2 }\,,\quad M_{2N+10}=\frac{\sqrt{q_1 q_2}}{M}\,.
    \label{eq:normal_Higgs}
\end{align}
Using the above Higgsing equation, and a further subtraction of the contribution from Goldstone bosons
\begin{align}
    Z_{\rm G.B.}&=\PE\left[\frac{\sqrt{q_1 q_2}}{(1-q_1)(1-q_2)(1-p)}(M+M^{-1} p)\sum_{f=1}^{2N+8}(M_f+M_f^{-1})\right]\,,
\end{align}
one can find the perturbative partition function (\ref{eq:perturbative_partition_function}) of $\sprm(N)$ from $\sprm(N+1)$.

\paragraph{Instanton partition function}
On the other hand, from the brane configuration Fig.\,\ref{fig:branes}, one can also study the instanton string corrections to the $\sprm(N)$ partition function, where the instanton strings are realized in terms of the D2-branes denoted as the red lines. The world-volume theory on a stack of $k$ D2 branes furnishes the $k$-th instanton string ADHM construction \cite{Kim:2014dza, Kim:2015fxa} in terms of 2d $\Ncal=(0,4)$ $\mathrm{O}(k)$ gauge theories with matter contents and interactions specified by the following quiver diagram
\begin{align}
    \raisebox{-.5\height}{
    \begin{tikzpicture}
	\tikzstyle{gauge} = [circle, draw,inner sep=3pt];
	\tikzstyle{flavour} = [regular polygon,regular polygon sides=4,inner sep=3pt, draw];
	\node (g1) [gauge,label=below:{$\orm(k)$}] {};
	\node (f1) [flavour,above of=g1, label=above:{$\sprm(N)$}] {};
	\node (f2) [flavour,right of=g1, label=right:{$\sorm(2N_f)$}] {};
	\draw (g1)--(f1);
	\draw[dashed] (g1)--(f2);
	\draw (g1) to [out=140,in=220,looseness=10] (g1);
	\node at (-1,0) {\small{sym}};
	\end{tikzpicture}
    } 
    \quad \text{with } N_f=2N+8,
    \label{eq:2d_quiver}
\end{align}
where the solid/dashed lines denote 2d hypermultiplets/Fermi multiplets, respectively. 

The $k$-th instanton string correction to the 6d partition function can be computed in terms of the elliptic genera of the 2d world-volume gauge theories\cite{Benini:2013nda, Benini:2013xpa}.
\begin{align}
 Z_{k}^{\rm 6d} = & \sum_{a} \frac{1}{|W_a| (2 \pi i )^r }\oint \prod_{i=1}^k \left(2\pi \eta^2 {\rm d}u_i 
\frac{{\vartheta}_1(2\epsilon_+)}{i\eta} \right)
 \prod_{e\in \mathrm{root}} \frac{\vartheta_1(e(u)) \vartheta_1(2\epsilon_+ +e(u))}{-\eta^2}  \notag \\
 & \times
 \prod_{\rho\in \mathrm{sym}} \frac{-\eta^2}{\vartheta_1(\epsilon_{1,2}+\rho(u) )} 
 \,\, \prod_{\mathclap{\rho\in \mathrm{fund}}} \,\, \left( \prod_{i=1}^N \prod_{f=1}^{2N+8}\frac{{\vartheta}_1(m_f +\rho(u))}{\eta^8{\vartheta}_1(\epsilon_+ +\rho(u) \pm
\alpha_i)}    \right)  \,,
\label{eq:pf_rough}
\end{align}
where 
$u$, $\alpha$, and $m$'s are the fugacities of the 2d gauge, hyper and fermi multiplets respectively, and we also denote $\epsilon_{\pm}\equiv\frac{1}{2}(\epsilon_1\pm\epsilon_2)$.  Here we use $a$ to label disconnected sectors of ${\rm O}(k)$ flat connections and $W_a$ is the Weyl group \cite{Kim:2014dza}. For illustration, we list here the elliptic genera of $\orm(2k)$ in the sector of trivial flat connection,
\begin{align}
Z_{2k,\,\mathrm{cont.}}^{\rm 6d}
=& \int \prod_{i=1}^k \diff u_i Z^{\rm 6d}_{k}(u)\equiv\,\,  \frac{1}{2^k k!}
 \int \prod_{i=1}^k \diff u_i \left( \frac{  \eta^3 {\vartheta}_1(2 
\epsilon_+)}{{\vartheta}_1( \epsilon_1) {\vartheta}_1( \epsilon_2)} \right)^k
\notag\\
&\, \times 
 \prod_{i=1}^k\frac{-\eta^2}{{\vartheta}_1(\pm 2u_i+\epsilon_{1,2})} \,\,\,\prod_{\mathclap{1\leq i< j\leq k}}
\quad \frac{{\vartheta}_1 (\pm u_i\pm u_j)  {\vartheta}_1 
(\pm u_i\pm u_j+2\epsilon_+)}{{\vartheta}_1(\pm u_i\pm u_j+\epsilon_1) 
{\vartheta}_1(\pm u_i\pm u_j+\epsilon_2)}\notag\\
&\, \times \prod_{i=1}^k \left(\frac{1}{\eta^8}\prod_{j=1}^{N}\frac{1}{{\vartheta}_1 (\pm\alpha_j\pm u_i+\epsilon_{+})}\prod_{l=1}^{2N+8}{\vartheta}_1(\pm u_i+m_l)\right)\,,
\label{eq:pf_cont}
\end{align}
where $2^k k!$ is the order of the Weyl group of $\orm(2k)$. 
One can check again that, using the Higgsing eq.\,(\ref{eq:normal_Higgs}), the contributions in eq.\,(\ref{eq:pf_cont}) from 6d gauge and hyper multiplets give
\begin{align}
\prod_{i=1}^k\frac{\vartheta_1(\pm u_i+m_{2N+10}) \vartheta_1(\pm u_i+m_{2N+9})}{\vartheta_1(\pm \alpha_N\pm u_i+\epsilon_+)}=1\,,
\end{align}
implying that the instanton string contribution for $\sprm(N+1)$ theory can be Higgsed to the $\sprm(N)$ one for continuous sector. It is also true for other discrete sectors that we will not present here for brevity. The instanton string corrections to the 6d partition function are obtained by summing up all these $k$-th elliptic genera,
\begin{align}
Z^{\rm 6d}_{\rm inst}=1+\sum_{k=1}^\infty \mathfrak q^k\, Z_k^{\rm 6d}=1+\sum_{k=1}^\infty \mathfrak q^k \left(Z_{k,\,{\rm dis.}}^{\rm 6d}+Z_{k,\,{\rm cont.}}^{\rm 6d}\right)\,,
\end{align}
where $\mathfrak q=e^{\phi_0}$ with $\phi_0$ being the tensor multiplet fugacity. 

Here for later use, we write down the result of the instanton partition function up to 2-instanton. One important but subtle point is that, the contributions in the integrands, say eq.\,(\ref{eq:pf_rough}) for example, are in fact from \emph{real} bosons or fermions in the 2d ADHM construction. Therefore each of them only contributes to a ``square root'' of $\vartheta_1$ function, see more details in \cite{Kim:2014dza, Chen:2021ivd}. Accordingly, the $``\vartheta_1"$ function therein would be appropriately understood as the product of two ``square root'' of $\vartheta_1$. It turns out to be important when one evaluates the instanton partition functions for the gauge fugacities $``u"$ taking values of discrete holonomies of $\orm(k)$, that we should have
\begin{align}
    \vartheta_1(u+z)\equiv\sqrt{\vartheta_1(u+z)\, \vartheta_1(-u+z)}\,,\quad {\rm for}\ \quad u\in\left\{0,\frac{1}{2},\frac{\tau+1}{2}, \frac{\tau}{2}\right\}\,,
\label{eq:square_root}
\end{align}
where the variable $``z"$ stands for flavor fugacities, Coulomb moduli and so on. 

With these preparations, the 1-instanton partition function is given by summing over 4 discrete holonomies of $\orm(1)\simeq \mathbb Z_2$,
\begin{align}
Z_1^{\rm 6d}=-\frac{1}{2\eta^6 {\vartheta}_1(\epsilon_{1,2})}\sum_{a=1}^4\frac{\prod_{f=1}^{2N+8}{\vartheta}_a(m_f)}{\prod_{i=1}^N{\vartheta}_a(\pm\alpha_i+\epsilon_+)}\,.
\label{eq:1-inst}
\end{align}
For the 2-instanton computation, notice that there is a continuous one and 6 discrete holonomies respect to $\orm(2)$. We sum over all these contributions and have
\begin{align}
Z_2^{\rm 6d}=\frac{1}{2}Z^{\rm 6d}_{2,\, {\rm cont.}}+\frac{1}{4}\sum_{a=2}^4 Z^{\rm 6d}_{2(a),\,{\rm dis.}}\,,
\label{eq:2-inst}
\end{align}
with
\begin{align}
Z_{2,\,{\rm cont.}}^{\rm 6d}\, = \, & 
  \frac{1}{{2}\eta^{12} 
  {\vartheta}_1(\epsilon_{1,2}) {\vartheta}_1(2\epsilon_1){\vartheta}_1(\epsilon_2-\epsilon_1)}
 \sum_{a=1}^4
 \left(
 \frac{\prod_{f=1}^{2N+8}
 {\vartheta}_a(m_f \pm {\epsilon_{1}}/{2} )
 }{\prod_{i=1}^N{\vartheta}_a(\pm\alpha_i+\epsilon_+\pm\epsilon_1/2)} 
  +(\epsilon_1\leftrightarrow\epsilon_2) \right)\notag\\*
 & \, +\sum_{i=1}^N\bigg(\frac{1}{\eta^{12}{\vartheta}_1(\epsilon_{1,2}){\vartheta}_1(2\alpha_i){\vartheta}_1(2\epsilon_+-2\alpha_i){\vartheta}_1(2\alpha_i-\epsilon_{1,2}){\vartheta}_1(-2\alpha_i+2\epsilon_++\epsilon_{1,2})}\notag\\*
 &\quad\quad \times \frac{\prod_{f=1}^{2N+8}{\vartheta}_1(\alpha_i\pm m_f-\epsilon_+)}{\prod_{j\neq i}^N{\vartheta}_1(\alpha_i\pm\alpha_j){\vartheta}_1(\alpha_i\pm\alpha_j-2\epsilon_+)}+(\alpha_i\rightarrow-\alpha_i) \bigg),
\end{align}
and
\begin{align}
Z^{\rm 6d}_{2(a),\,{\rm dis.}} \, = \, & \, 
 \frac{{\vartheta}_a(0) {\vartheta}_a(2\epsilon_+)}{\eta^{12} {\vartheta}_1(\epsilon_{1,2})^2\,
 {\vartheta}_a(\epsilon_{1,2})} 
 \bigg(  \frac{  \prod_{f=1}^{2N+8}  {\vartheta}_1(m_f ) {\vartheta}_a(m_f) }{\prod_{i=1}^N{\vartheta}_1(\pm\alpha_i+\epsilon_+){\vartheta}_a(\pm\alpha_i+\epsilon_+)} + \prod_{i=1}^2 \frac{  \prod_{f=1}^{2N+8}  {\vartheta}_{\sigma^i(a)}(m_f )  }{\prod_{i=1}^N{\vartheta}_{\sigma^i(a)}(\pm\alpha_i+\epsilon_+)}\bigg)\,,
\end{align}
where $\sigma=(234)$ is a permutation.

Overall, the full 6d $\sprm(N)$ partition function can be obtained by collecting both perturbative and instanton pieces together
\begin{align}
Z^{\sprm(N)}_{\rm 6d}=Z_{\rm pert} \cdot Z_{\rm inst}\,.
\end{align}
\vspace{2pt}

\subsection{Codimension two defect partition function}
Now we introduce a codimension two surface defect into the 6d $\sprm(N)$ theory. It can be done by turning on a spacetime dependent VEV to the meson operator of the to-be-Higgsed hypers in the $\sprm(N+1)$ theory. In the brane picture, one can interpret it as a D4 brane stretching between the NS5 and the moved D6 branes,
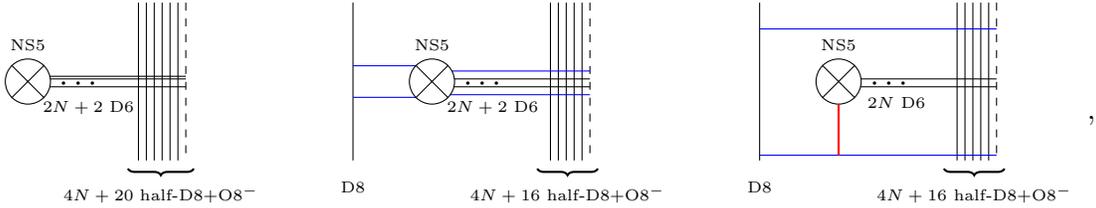
\begin{align}
\begin{gathered}
\begin{tikzpicture}[scale=0.70]
    \draw[dashed] (2,1.5)--(2,-1.5);
    \foreach \i in {1,...,6}
    {
    \draw (2-0.15*\i,1.5)--(2-0.15*\i,-1.5);
    }
    \draw (-1,0.1)--(2,0.1);
    \draw (-1,0.05)--(2,0.05);
    \node at (0,-0.05) {$\cdots$};
    \draw (-1,-0.1)--(2,-0.1);
    \ns{-1,0};
    \node at (-1,0.7) {\tiny{NS5}};
    \node at (0.15,-0.5) {\tiny{$2N+2$ D6}};
    \draw[decoration={brace,mirror,raise=23pt},decorate,thick](0.9,-0.5) -- node[below=25pt] {\tiny{$4N+20$ half-D8+O$8^-$}  } (2.15,-0.5);
    \end{tikzpicture}
\end{gathered}
\quad\quad
\begin{gathered}
\begin{tikzpicture}[scale=0.70]
    \draw[dashed] (2,1.5)--(2,-1.5);
    \foreach \i in {1,...,5}
    {
    \draw (2-0.15*\i,1.5)--(2-0.15*\i,-1.5);
    }
    \draw (-2.5,1.5)--(-2.5,-1.5);
    \draw[blue] (-1,0.3)--(-2.5,0.3);
    \draw[blue] (-1,-0.3)--(-2.5,-0.3);
    \draw[blue] (-1,0.2)--(2,0.2);
    \draw[blue] (-1,-0.25)--(2,-0.25);
    \draw (-1,0.05)--(2,0.05);
    \node at (0,-0.05) {$\cdots$};
    \draw (-1,-0.1)--(2,-0.1);
    \ns{-1,0};
    \node at (-1,0.7) {\tiny{NS5}};
    \node at (0.15,-0.5) {\tiny{$2N+2$ D6}};
    \node at (-2.5, -2) {\tiny D8};
    \draw[decoration={brace,mirror,raise=23pt},decorate,thick](1,-0.5) -- node[below=25pt] {\tiny{$4N+16$ half-D8+O$8^-$}  } (2.15,-0.5);
    \end{tikzpicture}
\end{gathered}
\quad\quad
\begin{gathered}
\begin{tikzpicture}[scale=0.70]
    \draw[dashed] (2,1.5)--(2,-1.5);
    \foreach \i in {1,...,5}
    {
    \draw (2-0.15*\i,1.5)--(2-0.15*\i,-1.5);
    }
    \draw (-2.5,1.5)--(-2.5,-1.5);
    \draw[blue] (-2.5,1)--(2,1);
    \draw[blue] (-2.5,-1.4)--(2,-1.4);
    \draw (-1,0.05)--(2,0.05);
    \node at (0,-0.05) {$\cdots$};
    \draw (-1,-0.1)--(2,-0.1);
    \ns{-1,0};
    \draw[red, thick] (-1,-.44)--(-1,-1.4);
    \node at (-1,0.7) {\tiny{NS5}};
    \node at (0.1,-0.4) {\tiny{$2N$ D6}};
    \draw[decoration={brace,mirror,raise=23pt},decorate,thick](1,-0.5) -- node[below=25pt] {\tiny{$4N+16$ half-D8+O$8^-$}  } (2.15,-0.5);
    \node at (-2.5, -2) {\tiny D8};
    \end{tikzpicture}
\end{gathered}
\,,\notag
\end{align}
where the red line denotes a D4 brane along directions of $x_{1, 2, 3, 4, 7, 8}$. When the D6 brane goes to infinity, the D4 brane becomes immobilized and serves as a codimension two defect. 

At the level of the partition function, the perturbative and instanton partition functions will both receive modifications. In the $\sprm(N+1)$ theory, we can now assign the following fugacities,
\begin{align}
 A_{N+1} : {M}{q_2}\,,\quad M_{2N+9} : M{q_2 \sqrt{q_1 q_2}}\,,\quad M_{2N+10} : \frac{\sqrt{q_1 q_2}}{M}\,,
\end{align}
or in exponents
\begin{align}
\alpha_{N+1}=-x+\epsilon_2\,,\ \ m_{2N+9}=-x+\epsilon_++\epsilon_2\,\ \ {\rm and}\ \ m_{2N+10}=x+\epsilon_+\,,
  \label{eq:defect_Higgs}
\end{align}
where we have introduced the defect parameter $X = e^x$ such that $M = X^{-1} $. 
It follows that the VEV of the meson $\mathcal M^{2N+9,2N+10}$ in eq.\,(\ref{eq:meson}) has fugacity
\begin{align}
\left\langle \mathcal M^{2N+9,2N+10}\right\rangle=q_2^{-1}=e^{-\epsilon_2}\,,
\label{eq:defect_meson}
\end{align}
which implies that it takes non-zero angular momentum with respect to the $\Omega$-background parameter $\epsilon_2$. We thus have introduced a vortex-like codimension-two defect localized on the space of $x_{4,5}$. In the rest of this subsection, we will use eq.\,(\ref{eq:defect_Higgs}) to compute the partition function in presence of a codimension two defect.

\subsubsection{Perturbative contribution}\label{sec:cod2_pert}
We now compute the perturbative partition function with a codimension two defect under the Nekrasov-Shatashvili limit, i.e. $\epsilon_2\rightarrow 0$ or equivalently $q_2\rightarrow 1$. Utilizing a similar method described in \cite{Chen:2021ivd} as well as eq.\,(\ref{eq:defect_Higgs}) to \eqref{eq:perturbative_partition_function}, we subtract it from $Z_{\rm G.B.}$ and taking $q_2\rightarrow 1$, we end up with the codimension two defect partition function of the $\sprm(N)$ theory
\begin{align}
    \frac{Z^{\rm{6d/4d}}_{\rm pert}(x)}{Z^{\rm{6d/4d}}_{\rm class}(x)}\, = \, & \PE\bigg[\frac{X^{-2}- p q X^{2} }{(1-q)(1- p)} +\sum_{i=1}^N\frac{ X^{-1} A_i - p q X A_i^{-1}   }{(1-q)(1- p)} \notag\\
    &
    \qquad -\sum_{i=1}^N\frac{ q X A_i - p X^{-1}A_i^{-1}  }{(1-q)(1- p)}
    - \sum_{f=1}^{2N+8}\frac{q^{\frac{1}{2}} \left(  X^{-1} M_f - X M_f^{-1} p \right) }{(1-q)(1- p)} 
 \bigg]\notag\\
     =& \prod_{i=1}^N\prod_{f=1}^{2N+8}\frac{\Gamma_{p,q}(X^{-2})\Gamma_{p,q}(X^{-1} A_i)}{\Gamma_{p, q}(q X A_i)\Gamma_{p,q}(q^{\frac{1}{2}}X^{-1}M_f)}\,,
\label{eq:pert_pf}
\end{align}
where $X = e^{x}$ is the defect parameter
and $\Gamma_{p, q}(z)$ is the elliptic gamma function
\begin{align}
   \Gamma_{p,q}(z) = \PE\bigg[\frac{z- p q/z}{(1-q)(1- p)}\bigg]\,.
\end{align}
Hereafter we will write $q$ to refer to  $q_1$.

Similarly, we can find the classical contributions to the codim two partition functions is
\begin{align}
Z_{\rm class}^{\rm 6d/4d}(x)=\exp\left\{-\frac{x}{2}+\frac{x}{\epsilon_1}\left(\sum_{i=1}^N\alpha_i-\frac{1}{2}\sum_{f=1}^{2N+8} m_f+\phi_0+\frac{\tau}{2}\right)+\frac{N+2}{2\epsilon_1}x^2\right\}\,.
\label{eq:class_pf}
\end{align}

\subsubsection{Instanton corrections}
For the instanton contribution, once again from eq.\,(\ref{eq:defect_Higgs}), we have
\begin{flalign}
Z^{\rm 4d}_{k,\,{\rm cont.}}(x)\equiv\prod_{i=1}^k\frac{\vartheta_1(\pm u_i+m_{2N+10})\vartheta_1(\pm u_i+m_{2N+9})}{\vartheta_1(\pm \alpha_N\pm u_i+\epsilon_+)}=\prod_{i=1}^k\frac{\vartheta_1(\pm u_i+x+\epsilon_+)}{\vartheta_1(\pm u_i+x+\epsilon_-)}\,, &&
\end{flalign}
in eq.\,(\ref{eq:pf_cont}), where we have shifted $x\rightarrow x-\frac{\epsilon_2}{2}$. Therefore we have
\begin{align}
Z_{2k,\,\mathrm{cont.}}^{{\rm 6d/4d}}
 = &\frac{1}{2^k k!}
 \int \prod_{i=1}^k \diff u_i \left( \frac{\eta^3 \vartheta_1(2 
\epsilon_+)}{\vartheta_1( \epsilon_1) \vartheta_1( \epsilon_2)} \right)^k  
\notag \\
&\, \times \prod_{i=1}^k\frac{-\eta^2}{\vartheta_1(\pm 2u_i+\epsilon_{1,2})}  \prod_{ 1\leq i< j\leq k} 
\frac{\vartheta_1 (\pm u_i\pm u_j)  \vartheta_1 
(\pm u_i\pm u_j+2\epsilon_+)}{\vartheta_1(\pm u_i\pm u_j+\epsilon_1) 
\vartheta_1(\pm u_i\pm u_j+\epsilon_2)} \notag\\
&\,\times \prod_{i=1}^k \left(\prod_{j=1}^{N}\frac{\eta^4}{\vartheta_1 (\pm\alpha_j\pm u_i+\epsilon_{+})}\prod_{l=1}^{2N+8}\frac{\vartheta_1(\pm u_i+m_l)}{-\eta^2}\right)
\prod_{i=1}^k\frac{\vartheta_1(\pm u_i+x+\epsilon_+)}{\vartheta_1(\pm u_i+x+\epsilon_-)}\notag\\
= & \int \prod_{i=1}^k \diff u_i\, Z^{\rm 6d}_{k}(u)\, Z_{k,\,{\rm cont.}}^{\rm 4d}(u, x)\,,
\label{eq:defect_pf_cont}
\end{align}
for the continuous sector of the instanton partition function with codim two defects. The computation is similar for the discontinuous sectors too. Practically, one has to be careful with the choices of poles in the integrand in presence of the defect contribution, e.g. $Z^{\rm 4d}_{k,\,{\rm cont.}}(x)$. Instead, we will directly calculate the JK-residues for the $\sprm(N+1)$ partition function without defects, and apply eq.\,(\ref{eq:defect_Higgs}) to the result therein. Once we obtain the defect partition function $Z_{\rm inst}^{\rm 6d/4d}(x;\epsilon_1,\epsilon_2)$, we define the normalized defect partition function $\Psi_{\rm inst}(x;\epsilon_1)$, when taking the NS-limit, as
\begin{align}
\Psi_{\rm inst}(x;\epsilon_1)\equiv \lim_{\epsilon_2\rightarrow 0}\frac{Z_{\rm inst}^{\rm 6d/4d}(x;\epsilon_1,\epsilon_2)}{Z_{\rm inst}^{\rm 6d}(\epsilon_1,\epsilon_2)}\,.
\label{eq:defect_pf_NS}
\end{align}
One will find that $\Psi_{\rm inst}(x;\epsilon_1)$ serves the wave function of the $\sprm(N)$ quantum curve in later sections.

To have a better sense of the instanton partition function with codim two defects, we here spell out 
$\Psi_{\rm inst}(x;\epsilon_1)$'s one and two-instanton order results. We first compute the refined defect partition function, and then take the NS-limit. For 1-instanton, one can directly apply eq.\,(\ref{eq:defect_Higgs}) to eq.\,(\ref{eq:1-inst}) and (\ref{eq:2-inst}), and have
\begin{align}
Z_1^{\rm 6d/4d}=-\frac{1}{2\eta^6\vartheta_1(\epsilon_{1,2})}\sum_a\frac{\prod_{f=1}^{2N+8}\vartheta_a(m_f)}{\prod_{i=1}^N\vartheta_a(\pm\alpha_i+\epsilon_+)}\, \frac{\vartheta_a(x+\epsilon_+)}{\vartheta_a(x+\epsilon_-)}\,.
\label{eq:1-inst_defect}
\end{align}
The two-instanton results have been put in Appendix \ref{B}.

Now we take the NS-limit to compute the normalized partition function with a codim two defect. From eq.\,(\ref{eq:defect_pf_NS}), up to 2-instanton order, we have
\begin{align}
\Psi_{\rm inst}(x;\epsilon_1)&\equiv 1+\mathfrak q\,\Psi_{1}(x;\epsilon_1)+\mathfrak q^2\,\Psi_2(x;\epsilon_1),
\end{align}
where 
\begin{align}
\Psi_{1}(x;\epsilon_1) &=\lim_{\epsilon_2\rightarrow 0}
\left(Z^{\rm 6d/4d}_1-Z^{\rm 6d}_1\right), \\*
\Psi_{2}(x;\epsilon_1) &= \lim_{\epsilon_2\rightarrow 0} \left( Z^{\rm 6d/4d}_2-Z^{\rm 6d}_2-Z^{\rm 6d}_1\left(Z^{\rm 6d/4d}_1-Z^{\rm 6d}_1\right) \right) .
\end{align}
Explicitly, we have the one-instanton result
\begin{align}
\Psi_{1}(x;\epsilon_1)=-\frac{1}{2\eta^6\vartheta_1(\epsilon_{1})\vartheta_1^\prime(0)}\sum_a\frac{\prod_{f=1}^{2N+8}\vartheta_a(m_f)}{\prod_{i=1}^N\vartheta_a(\pm\alpha_i+\frac{\epsilon_1}{2})}\frac{\vartheta^\prime_a\left(x+\frac{\epsilon_1}{2}\right)}{\vartheta_a\left(x+\frac{\epsilon_1}{2}\right)}\,,
\label{eq:1-inst_Psi}
\end{align}
and $\Psi_2(x;\epsilon_1)$ can be found in Appendix \ref{B}.

\subsection{Wilson surface defect}
In this subsection, we will discuss another important non-local supersymmetric observable, the Wilson surface defect as a codimension four defect in the 6d $\sprm(N)$ theory. 
One can introduce the codim four defects via a double Higgsing procedure from $\sprm(N+2)$ down to $\sprm(N)$ \cite{Kimura:2017auj, Chen:2021ivd}, as we will proceed below. 

In eq.\,(\ref{eq:mesons}), one may choose the operator $\mathcal M^{2N+9,2N+10}$ and $\widetilde{\mathcal M}_{2N+11,2N+12}$ to Higgs the $\sprm(N+2)$ theory. One assigns them to spacetime dependent VEVs as eq.\,(\ref{eq:defect_meson}) with fugacities
\begin{align}
\left[\left\langle \mathcal M^{2N+9,2N+10}\right\rangle\right]
=\left[\left\langle \widetilde{\mathcal M}_{2N+11,2N+12}\right\rangle\right]
=q_2^{-1} \equiv e^{-\epsilon_2}. 
\end{align}  
We thus have the following fugacities,
\begin{align}
 &A_{N+1}={M}{q_2}\,,\quad M_{2N+9}={M}{q_2 \,\sqrt{q_1 q_2}}\,,\quad M_{2N+10}=\frac{\sqrt{q_1 q_2}}{M},
\notag\\
&A_{N+2}=\frac{M^\prime}{q_2}\,,\quad M_{2N+11}=\frac{1}{\sqrt{q_1 q_2} M^\prime}\,,\quad M_{2N+12}=\frac{M^\prime}{q_2 \,\sqrt{q_1 q_2}}\,,
\end{align}
where we have used
\begin{align}
M\equiv e^{-\mu}\,\ \ {\rm and}\ \ M^{\prime}\equiv e^{-\nu}\,,
\end{align}
to parametrize the two codim two defects. To uniquely determine the Wilson surface defect, one needs to further require
\begin{align}
\mu+\epsilon_1=\nu-\epsilon_1\equiv x,
\label{eq:fine-tuning}
\end{align}
where $x$ is the fugacity of the Wilson defect. In terms of exponents, we have the following double Higgsing equation
\begin{align}
&\alpha_{N+1}=-x+2\epsilon_+\,,\ \ m_{2N+9}=-x+3\epsilon_+\,,\ \ m_{2N+10}=x+\epsilon_+-\epsilon_1,\notag\\
&\alpha_{N+2}=-x-2\epsilon_+\,,\ \ m_{2N+11}=x-\epsilon_++\epsilon_1\,,\ \ m_{2N+12}=-x-3\epsilon_+\,.
\label{eq:double_Higgs}
\end{align}

The double Higgsing can also be illustrated from a brane picture that the two codim two defects sitting on the NS5 brane can also be with each other. They then are free to move away, and there is an additional D2 brane stretching between the leaving D4 and NS5 branes. The D2 brane is terminated on another D4 brane spanning $x_{0,1,7,8,9}$, denoted as D$4^\prime$. Pulling the D$4^\prime$ brane back will annihilate the D2 brane, and finally the NS5-D6-D8/O$8^-$-D$4^\prime$ brane configuration gives the 6d/2d coupled system,
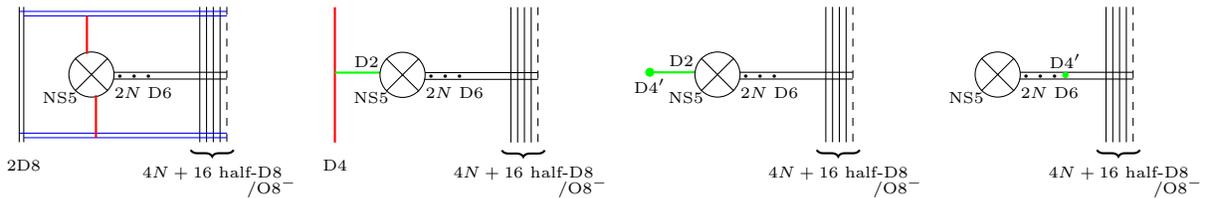
\begin{align}
\begin{gathered}
\begin{tikzpicture}[scale=0.60]
    \draw[dashed] (2,1.5)--(2,-1.5);
    \foreach \i in {1,...,4}
    {
    \draw (2-0.15*\i,1.5)--(2-0.15*\i,-1.5);
    }
    \draw (-2.5,1.5)--(-2.5,-1.5);
    \draw (-2.6,1.5)--(-2.6,-1.5);
    \draw[blue] (-2.6,1.4)--(2,1.4);
    \draw[blue] (-2.5,1.3)--(2,1.3);
    \draw[blue] (-2.5,-1.4)--(2,-1.4);
    \draw[blue] (-2.6,-1.3)--(2,-1.3);
    \draw (-1,0.05)--(2,0.05);
    \node at (0,-0.05) {$\cdots$};
    \draw (-1,-0.1)--(2,-0.1);
    \ns{-1,0};
    \draw[red, thick] (-0.9,-.46)--(-.9,-1.4);
    \draw[red, thick] (-1.1,.46)--(-1.1,1.3);
    \node at (-1.7, -0.5) {\tiny{NS5}};
    \node at (0.15,-0.4) {\tiny{$2N$ D6}};
    \draw[decoration={brace,mirror,raise=20pt},decorate,thick](1.2,-0.5) -- node[below=22pt] {\tiny{$4N+16$ half-D8}  } (2.15,-0.5);
    \node at (-2.5, -2) {\tiny 2D8};
    \node at (3, -2.50) {\tiny /O$8^-$};
    \end{tikzpicture}
\end{gathered}
\,
\begin{gathered}
\begin{tikzpicture}[scale=0.60]
    \draw[dashed] (2,1.5)--(2,-1.5);
    \foreach \i in {1,...,4}
    {
    \draw (2-0.15*\i,1.5)--(2-0.15*\i,-1.5);
    }
    \draw[red, thick] (-2.5,1.5)--(-2.5,-1.5);
    \draw[green, thick] (-2.5, 0.05)--(-1,0.05);
    \draw (-1,0.05)--(2,0.05);
    \node at (0,-0.05) {$\cdots$};
    \draw (-1,-0.1)--(2,-0.1);
    \ns{-1,0};
    \node at (-1.7, -0.5) {\tiny{NS5}};
    \node at (-1.8, +0.3) {\tiny{D2}};
    \node at (0.15,-0.4) {\tiny{$2N$ D6}};
    \draw[decoration={brace,mirror,raise=20pt},decorate,thick](1.2,-0.5) -- node[below=22pt] {\tiny{$4N+16$ half-D8}  } (2.15,-0.5);
    \node at (-2.5, -2) {\tiny D4};  
    \node at (3, -2.50) {\tiny /O$8^-$};
    \end{tikzpicture}
\end{gathered}
\,
\begin{gathered}
\begin{tikzpicture}[scale=0.60]
    \draw[dashed] (2,1.5)--(2,-1.5);
    \foreach \i in {1,...,4}
    {
    \draw (2-0.15*\i,1.5)--(2-0.15*\i,-1.5);
    }
    \draw[green, thick] (-2.5, 0.05)--(-1,0.05);
    \draw (-1,0.05)--(2,0.05);
    \node at (0,-0.05) {$\cdots$};
    \draw (-1,-0.1)--(2,-0.1);
    \ns{-1,0};
    \node at (-1.7, -0.5) {\tiny{NS5}};
    \node at (-1.8, +0.3) {\tiny{D2}};
    \node at (0.15,-0.4) {\tiny{$2N$} D6};
    \draw[decoration={brace,mirror,raise=20pt},decorate,thick](1.2,-0.5) -- node[below=22pt] {\tiny{$4N+16$ half-D8}  } (2.15,-0.5);
    \fill[green] (-2.5, 0.05) circle (.1);
    \node at (-2.5, -0.25) {\tiny D$4^\prime$};
    \node at (3, -2.50) {\tiny /O$8^-$};
    \end{tikzpicture}
\end{gathered}
\,
\begin{gathered}
\begin{tikzpicture}[scale=0.60]
    \draw[dashed] (2,1.5)--(2,-1.5);
    \foreach \i in {1,...,4}
    {
    \draw (2-0.15*\i,1.5)--(2-0.15*\i,-1.5);
    }
    \draw (-1,0.05)--(2,0.05);
    \node at (0,-0.05) {$\cdots$};
    \draw (-1,-0.1)--(2,-0.1);
    \ns{-1,0};
    \node at (-1.7, -0.5) {\tiny{NS5}};
    \node at (0.15,-0.4) {\tiny{$2N$} D6};
    \draw[decoration={brace,mirror,raise=20pt},decorate,thick](1.2,-0.5) -- node[below=22pt] {\tiny{$4N+16$ half-D8}  } (2.15,-0.5);
    \fill[green] (0.5, -0.01) circle (.07);
    \node at (0.5, 0.3) {\tiny D$4^\prime$};
    \node at (3, -2.50) {\tiny /O$8^-$};
    \end{tikzpicture}
\end{gathered}
\notag
\end{align}

\paragraph{Instanton corrections}
Applying eq.\,(\ref{eq:double_Higgs}) to eq.\,(\ref{eq:pf_cont}), we have an additional term inserted in the instanton partition function,
\begin{align}
Z^{\rm 2d}_{k,\,{\rm cont.}}(x)\equiv\prod_{i=1}^k\frac{\vartheta_1(\pm u_i+x\pm\epsilon_-)}{\vartheta_1(\pm u_i+x\pm\epsilon_+)}\,,
\end{align}
regarded as the contribution from the Wilson surface defect. Therefore for the continuous sector, we have
\begin{align}
W^{\orm(2k)}_{2k,\,{\rm cont.}}(x)=\int \prod_{i=1}^k \diff u_i\, Z^{\rm 6d}_{k}(u) Z_{k,\,{\rm cont.}}^{\rm 2d}(u, x)\,. 
\end{align}

We will write down the Wilson surface defect up to 2-instanton. Applying eq.\,(\ref{eq:double_Higgs}) to (\ref{eq:1-inst}) and (\ref{eq:2-inst}), we obtain the one-instanton contribution
\begin{align}
W_1^{\rm 6d/2d}(x;  \epsilon_{1} , \epsilon_{2})=-\frac{1}{2\eta^6\vartheta_1(\epsilon_{1,2})}\sum_a\frac{\prod_{f=1}^{2N+8}\vartheta_a(m_f)}{\prod_{i=1}^N\vartheta_a(\pm\alpha_i+\epsilon_+)} \frac{\vartheta_a(x\pm\epsilon_-)}{\vartheta_a(x\pm\epsilon_+)}\,.
\label{eq:1-inst_Wilson}
\end{align}
The two-instanton contribution $W_2^{\rm 6d/2d}(x; \epsilon_1, \epsilon_2)$ can be found in Appendix \ref{B}.

Putting together the full-instanton contributions, we will get the Wilson surface defect $W^{\rm 6d/2d}_{\rm inst}(x;\epsilon_1,\epsilon_2)$. After taking NS-limit, we also define the normalized Wilson surface defect as
\begin{align}
\chi_{\rm inst}(x;\epsilon_1) & : =  \lim_{\epsilon_2\rightarrow 0}\frac{W^{\rm 6d/2d}_{\rm inst}(x;\epsilon_1,\epsilon_2)}{Z^{\rm 6d}_{\rm inst}(\epsilon_1,\epsilon_2)} \notag \\
& = 1+\mathfrak q\, \chi_{1}(x;\epsilon_1)+\mathfrak q^2\chi_2(x;\epsilon_1)+\mathcal O(\mathfrak q^3)\,,
\end{align}
In particular, the one-instanton contribution is given by
\begin{align}
\chi_1(x;\epsilon_1) = & \lim_{\epsilon_2\rightarrow 0}\left(W^{6d/2d}_1-Z^{6d}_1\right) \notag\\*
= & 
-\frac{1}{2\eta^{6}\vartheta_1(\epsilon_{1})\vartheta_1^\prime(0)}\sum_{a=1}^4 \frac{\prod_{f=1}^{2N+8}\vartheta_a(m_f)}{\prod_{i=1}^N\vartheta_a\left(\pm\alpha_i+\frac{\epsilon_1}{2}\right)}\vartheta_a^\Delta(x)\,,
\label{eq:1-inst_chi}
\end{align}
where $\vartheta_a^\Delta(x)$ is defined as (\ref{eq:thetadelta}). 
The two-instanton contribution $\chi_2(x; \epsilon_1)$ is given in Appendix \ref{B}.

\subsubsection{Perturbative contribution and poles in Wilson surface defect}

In the first part of this subsection, we write down the perturbative part of the Wilson surface defect partition function. 
\begin{align}
W_{\rm pert}=\mathfrak q^{-1}\prod_{i=1}^N\frac{\vartheta_1(x\pm \alpha_i)}{(i\eta)^2}=\mathfrak q^{-1}\prod_{i=1}^N\theta_1(x\pm\alpha_i)\,,
\label{eq:pert_Wilson}
\end{align}
where the factor $\mathfrak q^{-1}$ comes from the Green-Schwarz contribution of $\sprm(N+2)$ theory with the tuning of parameters in the double Higgsing equation \eqref{eq:double_Higgs}.
The second part in \eqref{eq:pert_Wilson} comes from open string degrees of freedom between the D$4^\prime$-D6 branes. Recall that the Wilson surface and the gauge multiplets are introduced by the D$4^\prime$ and D6 branes respectively in the 6d $\sprm(N)$ theory. The 2d fermions of the D$4^\prime$-D6 combine with bosons of the D2-D6 and fermions of the D2-NS5-D6 form the $\mathcal{N} =(0, 4)$ invariant superpotential in the 2d GLSM \cite{Agarwal:2018tso}.

In the rest of this subsection, we examine the pole structures of the Wilson surface defect, 
\begin{align}
\chi(x;\epsilon_1)\equiv W_{\rm pert}(x;\epsilon_1) \chi_{\rm inst}(x;\epsilon_1).
\end{align}
Firstly it is worth distinguishing the difference between the Wilson surface expectation values and the Wilson surface defect we have computed. This point is easily appreciated when one compactifies the 6d theory onto a circle and reduces it to a 5d KK theory. In this picture, all 6d tensors and gauge moduli turn out to be 5d gauge moduli. Therefore we can equivalently compute the 5d Wilson loop expectations with respect to these gauge multiplets. The 5d Wilson loop expectations are independent of the defect parameter ``$x$'', but rather some ($q$-deformed) characters with respect to the gauge groups. On the other hand, the Wilson surface defect, or the Wilson loop defect in the 5d perspective, is indeed $x$-dependent, and thus has to be regarded as a generating function of the Wilson surface expectations, i.e.
\begin{align}
\chi(x;\epsilon_1)=\sum_{n}W_n(\epsilon_1)\cdot\vartheta^{[n]}(x)\,,
\label{eq:decomposition}
\end{align}
where $W_n(\epsilon_1)$ is the $q$-deformed Wilson surface expectations, and $\vartheta^{[n]}(x)$ are some basis in terms of elliptic theta functions with a certain degree.

In \cite{Chen:2021rek}, for the 6d $\sorm(N)$ gauge theory, the above decomposition has been checked to hold up to three-instanton orders. However, for the $\sprm(N)$ theory, one subtlety occurs here. A Coulomb parameter independent part of the one-instanton correction contains nontrivial poles over the defect parameter $x$, which should be regarded as a part of the quantum curve. Such phenomenon happens for E-string theory when we focus on a trivial gauge group $\sprm(0)$, the one-instanton correction has been identified as the 4-theta potential of the van Diejen integrable system \cite{Chen:2021ivd}. Therefore, in the $\sprm(N)$ theory, after the subtraction of the $x$ singular part $\mathcal{V}(x;\epsilon_1,m_i)$ from the one-instanton partition function, the remaining holomorphic part shall again have the structure \eqref{eq:decomposition}, that is
\begin{align}
\chi(x;\epsilon_1)=\mathcal{V}(x)+\sum_{n}W_n(\epsilon_1)\cdot\vartheta^{[n]}(x)\,.
\label{eq:decomposition_sp}
\end{align}
We expect that $W_n(\epsilon_1)$ is the expectation value of the Wilson surface operator, which can be alternatively calculated in the 5d KK theory as the Wilson loop expectation value. A detail discussion about the decomposition \eqref{eq:decomposition_sp} can be found in Section \ref{sec:spcurve}

We further want to remark that, in the E-string and M-string case, the Wilson surface depends only on the Coulomb parameter coming from the tensor multiplet, for there is no gauge multiplet in the 6d theory, and the 2d fermions raised between D$4^\prime$-D6 should be decoupled in the IR. In contrast, for the 6d theories dressed with gauge multiplets, the 2d fermions from D$4^\prime$-D6 turn out to be important. In practice, they remove all poles of defect parameter raised from the gauge fugacities in the Wilson surface defect.

\subsection{$\sprm(N)$ quantum curve}\label{sec:spcurve}
Since the two important ingredients, the codim two and four defects, have been computed, we are ready to write down the $\sprm(N)$ quantum curve to connect them. The quantum curve of the simplest $\sprm(0)$ theory, or say the E-string, has been derived in \cite{Chen:2021ivd} from a path integral approach. It implies that, under the NS-limit, the E-string curve is dominated only by the continuous sector of its 2d $\orm(k)$ elliptic genera. One can perform a similar path integral analysis for the general $\sprm(N)$ case, from which we propose their quantized Seiberg-Witten curve for the instanton part:
\begin{align}
\mathcal D_{{\rm inst}}\Psi_{\rm inst}(x;\epsilon_1)=\chi_{\rm inst}(x;\epsilon_1)\,\Psi_{\rm inst}(x;\epsilon_1)\,,
\label{eq:SpN_curve_inst}
\end{align}
with
\begin{align}
\mathcal D_{{\rm inst}}&\equiv Y+\frac{\mathfrak q^2 }{\eta^{12}\vartheta_1(2x)\vartheta_1(2x+\epsilon_1)^2\vartheta_1(2x+2\epsilon_1)}\frac{\prod_{f=1}^{2N+8}\vartheta_1\left(x\pm m_f+\frac{\epsilon_1}{2}\right)}{\prod_{i=1}^N\vartheta_1(x\pm\alpha_i)\vartheta_1(x\pm\alpha_i+\epsilon_1)}\cdot Y^{-1}\,,
\notag\\\notag\\
\end{align}
where $Y$ is the difference operator satisfying
\begin{align}
Y X=Y e^x=e^{-\epsilon_1}X Y\,,
\end{align}
i.e. it shifts $x$ to $x-\epsilon_1$. Obviously, the $\sprm(N)$ curve includes the E-string curve as a special case with $N=0$. 

We can recast eq.\,(\ref{eq:SpN_curve_inst}) as
\begin{align}
\chi_{\rm inst}(x;\epsilon_1)=\frac{\mathcal D_{\rm inst} \Psi_{\rm inst}(x;\epsilon_1)}{\Psi_{\rm inst}(x;\epsilon_1)}\,.
\end{align}
then one can use eq.\,(\ref{eq:1-inst_Psi}), (\ref{eq:2-inst_Psi}), (\ref{eq:1-inst_chi}) and (\ref{eq:2-inst_chi}) to honestly check the curve (\ref{eq:SpN_curve_inst}) is satisfied up to the 2-instanton order.

Now we collect the perturbative contributions to the quantum curve. From eq.\,(\ref{eq:pert_pf}) and (\ref{eq:class_pf}), perturbative partition function spells as
\begin{align}
 \Psi_{\rm pert}(x)\equiv{Z^{\rm{6d/4d}}_{\rm pert}(x)}=Z_{\rm class}^{\rm 6d/4d}(x) \prod_{i=1}^N\prod_{f=1}^{2N+8}\frac{\Gamma_{p, q}(X^{-2})\Gamma_{p, q}(X^{-1} A_i)}{\Gamma_{p, q}(q X A_i)\Gamma_{p, q}(q^{\frac{1}{2}}X^{-1}M_f)},
\end{align}
with
\begin{align}
Z_{\rm class}^{\rm 6d/4d}(x)=\exp\left\{-\frac{x}{2}+\frac{x}{\epsilon_1}\left(\sum_{i=1}^N\alpha_i-\frac{1}{2}\sum_{f=1}^{2N+8} m_f+\phi_0+\frac{\tau}{2}\right)+\frac{N+2}{2\epsilon_1}x^2\right\}.
\end{align}
Acting the shift operator $Y$, one can find
\begin{align}
    Y Z^{\rm 6d/4d}_{\rm class}(x)=\left(\frac{q^{\frac{N+3}{2}}\prod_{f=1}^{2N+8}M_f^{\frac{1}{2}}}{p^{\frac{1}{2}} X^{N+2}\prod_{i=1}^N A_i}\mathfrak{q}^{-1}\right) Z^{\rm 6d/4d}_{\rm class}(x)\,,
    \label{eq:shift_class}
\end{align}
and
\begin{align}
    &Y \left(\prod_{i=1}^N\prod_{f=1}^{2N+8}\frac{\Gamma_{p,q}(X^{-2})\Gamma_{p,q}(X^{-1} A_i)}{\Gamma_{p,q}(pXA_i)\Gamma_{p,q}(p^{\frac{1}{2}}X^{-1}M_f)}\right) \notag\\
&=\left(\prod_{i=1}^N\prod_{f=1}^{2N+8}\frac{\left[q X^{-2}\right]\left[X^{-2}\right]\left[A_iX^{\pm 1}\right]}{\left[q^{\frac{1}{2}} M_f X^{-1}\right]}\right) \prod_{i=1}^N\prod_{f=1}^{2N+8}\frac{\Gamma_{p,q}(X^{-2})\Gamma_{p,q}(X^{-1} A_i)}{\Gamma_{p,q}(q X A_i)\Gamma_{p,q}(q^{\frac{1}{2}} X^{-1} M_f)}\,,
\label{eq:shift_pert}
\end{align}
where we have defined the $p$-theta function
\begin{align}
    [X]\equiv\prod_{k=0}^\infty(1-X p^k )(1-X^{-1}p^{k+1})=X^{\frac{1}{2}}p^{-\frac{1}{12}}\frac{\vartheta_1(x)}{i\eta}\,,
\label{eq:Qtheta}
\end{align}
and used the property of the elliptic Gamma function
\begin{align}
    \Gamma_{p,q}(pX)=[X]\Gamma_{p,q}(X)\,, \qquad {\rm or}\qquad \Gamma_{p,q}(p^{-1}X)=\frac{\Gamma_{p,q}(X)}{[p^{-1}X]}\,.
\end{align}
Therefore overall, combining eq.\,(\ref{eq:shift_class}) and (\ref{eq:shift_pert}), and further using eq.\,(\ref{eq:Qtheta}), we arrive at
\begin{align}
Y \Psi_{\rm pert}(x)=\left(\mathfrak q^{-1}\frac{\vartheta_1(2x-\epsilon_1)\vartheta_1(2x)\prod_{i=1}^N\vartheta_1(\pm x+\alpha_i)}{-\eta^{-6}\prod_{f=1}^{2N+8}\vartheta_1\left(x-m_f-\frac{\epsilon_1}{2}\right)}\right) \Psi_{\rm pert}(x),
\end{align}
or equivalently,
\begin{align}
Y^{-1} \Psi_{\rm pert}(x)=\left(\mathfrak q\, \frac{-\eta^{-6}\prod_{f=1}^{2N+8}\vartheta_1\left(x-m_f+\frac{\epsilon_1}{2}\right)}{\vartheta_1(2x+\epsilon_1)\vartheta_1(2x+2\epsilon_1)\prod_{i}^N\vartheta_1(\pm (x+\epsilon_1)+\alpha_i)}\right) \Psi_{\rm pert}(x).
\end{align}
Now we look at the full codimension two defect partition function
\begin{align}
\Psi(x;\epsilon_1)\equiv\Psi_{\rm pert}(x) \Psi_{\rm inst}(x;\epsilon_1)\,.
\end{align}
Assembling to the difference equation satisfied by the instanton piece \eqref{eq:SpN_curve_inst}, we have 
\begin{align}
\mathcal D_{\rm full} \Psi(x;\epsilon_1)=\chi_{\rm inst}(x;\epsilon_1) \Psi(x;\epsilon_1),
\end{align}
with
\begin{align}
\mathcal D_{\rm full}\equiv&\,\mathfrak q\,\frac{-\eta^{-6}\prod_{f=1}^{2N+8}\vartheta_1\left(x-m_f-\frac{\epsilon_1}{2}\right)}{\vartheta_1(2x-\epsilon_1)\vartheta_1(2x)\prod_{i=1}^N\vartheta_1(\pm x+\alpha_i)} Y\notag\\
&+\mathfrak q\,\frac{-\eta^{-6}\prod_{f=1}^{2N+8}\vartheta_1\left(x+m_f+\frac{\epsilon_1}{2}\right)}{\vartheta_1(2x)\vartheta_1(2x+\epsilon_1)\prod_{i=1}^N\vartheta_1(\pm x+\alpha_i)} Y^{-1}\,.
\end{align}
Notice that, because both $\vartheta_a$ and $\eta$ have modular weight $\frac{1}{2}$, the difference operator $\mathcal D_{\rm full}$ thus has modular weight $0$.  For later convenience, we introduce weight $0$ theta function
\begin{align}
\theta_a(x)\equiv\frac{\vartheta_a(x)}{i\eta}\,,
\end{align}
to simplify our expressions. Recall further from eq.\,(\ref{eq:pert_Wilson}) that
\begin{align}
W_{\rm pert}(x)=  \mathfrak q^{-1} \prod_{i=1}^N\frac{\vartheta_1(x\pm\alpha_i)}{(i\eta)^2}= \mathfrak q^{-1} \prod_{i=1}^N{\theta_1(x\pm\alpha_i)}\,.
\end{align}
We therefore multiply the difference equation by $ W_{\rm pert}(x)$, and rewrite it as\footnote{We have re-defined the wave-function by an additional phase factor $e^{N\pi i\frac{x}{\epsilon_1}}$, so that $e^{-N\pi i\frac{x}{\epsilon_1}} Y\cdot e^{N\pi i\frac{x}{\epsilon_1}}=(-1)^N$, to absorb the unwanted factor $(-1)^N$ in $\mathcal D_{\sprm(N)}$.}
\begin{align} 
\mathcal D_{\sprm(N)} \Psi(x)\equiv\left(V(x) Y+V(-x)  Y^{-1} + \chi(x)\right) \Psi(x)=0\,,
\label{eq:SpN_curve}
\end{align}
where we have defined
\begin{align}
V(x)\equiv W_{\rm pert}(x)\cdot\left(\mathfrak q\,\frac{-\eta^{-6}\prod_{f=1}^{2N+8}\vartheta_1\left(x-m_f-\frac{\epsilon_1}{2}\right)}{\vartheta_1(2x-\epsilon_1)\vartheta_1(2x)\prod_{i=1}^N\vartheta_1(x\pm\alpha_i)}\right)=
\frac{\prod_{f=1}^{2N+8}\theta_1\left(x-m_f-\frac{\epsilon_1}{2}\right)}{\theta_1(2x)\theta_1(2x-\epsilon_1)}\,,
\label{eq:SpN_curve_shift}
\end{align}
and
\begin{align}
\chi(x)&=- W_{\rm pert}\times\chi_{\rm inst}(x;\,\epsilon_1)\notag\\
&=\frac{1}{2\theta_1(\epsilon_{1})\theta_1^\prime(0)}\sum_{a=1}^4\prod_{i=1}^{N}\prod_{f=1}^{2N+8}\frac{\theta_1(x\pm\alpha_i)\theta_a(m_f)}{\theta_a(\pm\alpha_i+\epsilon_1/2)}\left(\frac{\theta^\prime_a\left(x-\frac{\epsilon_1}{2}\right)}{\theta_a\left(x-\frac{\epsilon_1}{2}\right)}-\frac{\theta^\prime_a\left(x+\frac{\epsilon_1}{2}\right)}{\theta_a\left(x+\frac{\epsilon_1}{2}\right)} \right)  + \mathcal{E}(x)\,,
\end{align}
is the perturbative and one-instanton correction to the Wilson surface defect. $\mathcal E(x)$ is the rest part of the Wilson surface defect. Note that we can express $\mathcal D_{\sprm(N)}$ using $\vartheta_a$ instead of $\theta_a$ if we multiply by an overall $\eta^{2N+6}$.  

We claim that $\chi(x)$ can be further rewritten into the following form
\begin{align}
    \chi(x) = \mathcal{V}(x) + \sum_{k=0}^N \mathcal{W}_k(\alpha_i) \, v_k(x).
\end{align}
where $v_k(x)$ for $k = 0, 1, \cdots, N$ is a basis for the even holomorphic theta functions of degree $2N$. Here $\mathcal{V}(x)$ is the singular part given by meromorphic theta functions in $x$ of degree $2N$ and it is independent of $\alpha_i$, and the rest of parts are regular with the coefficients $\mathcal{W}_k(\alpha_i)$ giving the vacuum expectation values of Wilson surfaces.

For this purpose, we will use two sets of useful theta/elliptic function identities summarized in Appendix \ref{A}. We can find\footnote{The singular part of the E-string curve presented in \cite{Chen:2021ivd} can be written in the same form if one does the same operation here.}
\begin{align}
    \mathcal{V}(x) = \frac{1}{2} \sum_{b=1}^{4} \frac{\prod_{f=1}^{2N+8} \theta_b(m_f) \prod_{i=0}^{N} (-1)^{e_{i,b}} \frac{\theta_{a_i}(x)^2}{\theta_{c_i}(\frac{\epsilon_1}{2})^2} }{\theta_b(x \pm \frac{ \epsilon_1}{2})},
\end{align}
with 
\begin{align}
    e_{i,b} = \begin{cases}
        \alpha_{a_0, b}, &  \text{for } i =0, \\
        \beta_{c_i, b},  & \text{for } i \neq 0 .
    \end{cases}
\end{align}
For the regular part, we can apply the identities \ref{squares} repeatedly until it can be expressed as an expansion using a chosen basis, such as $v_k(x) = \theta_a(x)^{2k} \theta_b(x)^{2N-2k}$, where $a \neq b$.. 

In the following, we explicitly present the $N=1$ case to illustrate the result. We will simply write $\alpha_1$ as $\alpha$. We will use the basis with $v_1(x)= \theta_1(x)^2$ and $v_2(x) = \theta_4(x)^2$. The singular part is given by
\begin{align}
    \mathcal{V}(x) & =  \sum_{b=1}^4 (-1)^{\delta_{b,3}}  \frac{\prod_{f=1}^{2N+8} \theta_b(m_f)} {2\theta_c(\frac{\epsilon_1}{2})^2 \theta_{c'}(\frac{\epsilon_1}{2})^2} \frac{\theta_a(x)^2 \theta_{a'}(x)^2}{\theta_b(x \pm \frac{\epsilon_1}{2})} \notag \\
    & = \frac{\prod_{f=1}^{2N+8} \theta_1(m_f)} {2\theta_4(\frac{\epsilon_1}{2})^4} \frac{\theta_4(x)^4}{\theta_1(x \pm \frac{\epsilon_1}{2})} + \sum_{b=2}^4 (-1)^{\delta_{b,3}}  \frac{\prod_{f=1}^{2N+8} \theta_b(m_f)} {2\theta_b(\frac{\epsilon_1}{2})^4} \frac{\theta_1(x)^4}{\theta_b(x \pm \frac{\epsilon_1}{2})},
\end{align}
where we made the choice such that $a = a_1 = 4$, $c = c_1 =4$ for $b=1$, and $a = a_1 = 1$, $c = c_1 =b$ for $b \neq 1$. 

We also have the one-instanton contribution
\begin{align}
   \mathcal{W}_1 & = - \frac{\prod_{f=1}^{2N+8} \theta_1(m_f)} {\theta_4(\frac{\epsilon_1}{2})^4} \frac{\theta_4(\alpha)^2}{\theta_1(\pm \alpha +\frac{\epsilon_1}{2})} \frac{\theta_4'(\frac{\epsilon_1}{2})\theta_4(\frac{\epsilon_1}{2})^3}{\theta_1(\epsilon) \theta_1'(0) \theta_4(0)^2} \notag\\
    & \quad + \sum_{b=2}^4 \frac{\prod_{f=1}^{2N+8} \theta_b(m_f)} {2\theta_b(\frac{\epsilon_1}{2})^4} \Big ( - 2\frac{\theta_b'(\frac{\epsilon_1}{2})\theta_b(\frac{\epsilon_1}{2})}{\vartheta_1(\epsilon) \theta_1'(0)} - (-1)^{\delta_{b,3}} \frac{\theta_1(\alpha)^2}{\theta_b(\pm \alpha + \frac{\epsilon_1}{2})} 
    \notag\\
    & \qquad 
    + 2 \frac{\theta_1(\alpha)^2}{\theta_b(\pm \alpha +\frac{\epsilon_1}{2})} \frac{\theta_b'(\frac{\epsilon_1}{2})\theta_b(\frac{\epsilon_1}{2}) \theta_c(\frac{\epsilon_1}{2})^2}{\theta_1(\epsilon) \theta_1'(0) \theta_4(0)^2}\Big), \\
\mathcal{W}_2 &= \frac{\prod_{f=1}^{2N+8} \theta_1(m_f)} {2\theta_4(\frac{\epsilon_1}{2})^4} \Big(- 2\frac{\theta_4'(\frac{\epsilon_1}{2})\theta_4(\frac{\epsilon_1}{2})}{\theta_1(\epsilon) \theta_1'(0)} + \frac{\theta_4(\alpha)^2}{\theta_1(\pm \alpha + \frac{\epsilon_1}{2})} \notag\\
    & \quad +  2 \frac{\theta_4(\alpha)^2}{\theta_1(\pm \alpha +\frac{\epsilon_1}{2})} \frac{\theta_4'(\frac{\epsilon_1}{2})\theta_4(\frac{\epsilon_1}{2}) \theta_1(\frac{\epsilon_1}{2})^2}{\theta_1(\epsilon) \theta_1'(0) \theta_4(0)^2}\Big) \notag\\
    & - \sum_{b=2}^4 \frac{\prod_{f=1}^{2N+8} \theta_b(m_f)} {\theta_b(\frac{\epsilon_1}{2})^4} \frac{\theta_1(\alpha)^2}{\theta_b(\pm \alpha +\frac{\epsilon_1}{2})} \frac{\theta_b'(\frac{\epsilon_1}{2})\theta_b(\frac{\epsilon_1}{2})^3}{\theta_1(\epsilon) \theta_1'(0) \theta_4(0)^2} ,
\end{align}
where the supercript $c$ appered in $\mathcal{W}_1$ satisfies $\omega_c = \omega_b + \omega_4$ (mod $\mathbb{Z} + \tau \mathbb{Z}$).
    
\section{\boldmath $\sprm(N)$ Quantum Curve as An Elliptic Garnier System}
\label{Garnier}
\subsection{Elliptic Garnier system}

In the proceeding section, we obtained the elliptic difference equation (\ref{eq:SpN_curve}) which quantizes the Seiberg-Witten curves of the 4d KK theory that come from 6d D-type minimal conformal theory on the torus. It is desirable to identify explicit integrable models whose phase spaces describe the Coulomb branches of our theories. In case such integrable models admit Lax representations, we can obtain the spectral curve as well as its quantization. Here the quantized spectral curve will be the analog of Opers that appear in the Hitchin moduli spaces \cite{Nekrasov:2011bc}. In this paper, we claim that the underlying integrable models are given by the elliptic Garnier systems which has been studied in the math literature \cite{OrmerodRains17,Yamada17}. There are two different Lax representations provided in these papers and their equivalence is not easy to establish. As we will see, the quantum curve derived from the Lax presentation given in Yamada's paper matches exactly with ours, following the method in \cite{MR4120359}. 

Now we follow \cite{MR4120359} to rewrite the Lax equation of the elliptic Garnier system in the formalism of quantum curves we derived in previous sections. Before proceeding, let us first clarify the notations. The torus moduli parameter $\tau$, $\Omega$-background deformation $\epsilon_1$, defect parameter $x$, and mass parameter $m_f$ in the context of Garnier models are given by
\begin{align}
\tau=- i a_+\,,\quad\quad \epsilon_1= - 2\pi a_-\,,\quad\quad x= \log z + \pi (a_+ + a_-)\,,\quad\quad m_f=-2\pi \gamma_f,
\end{align}
or in exponential form
\begin{align}
& p\equiv e^{2 \pi i \tau}=e^{-2\pi a_+}\,,\qquad q\equiv e^{\epsilon_1}=e^{ - 2 \pi a_-}\,,\notag\\
&z=e^{(x-\pi(a_+ + a_-))}=X\sqrt{p q}\,,\qquad M_f=e^{m_f}=e^{-2\pi\gamma_f}\,.
\end{align}
In addition, we introduce the following function $R_p(X)$ for convenience
\begin{align}
    R_p(X)\equiv\prod_{n=1}^{\infty}(1-X p^{n-\frac{1}{2}})(1-X^{-1} p^{n-\frac{1}{2}})=\left[\sqrt{p}X\right]\,,
\end{align}
where $[\ \cdot\ ]$ is the $p$-theta function defined in eq.\,(\ref{eq:Qtheta}).

\paragraph{Shift part}
Now we follow \cite{MR4120359} to spell out the Lax equation in the Garnier models,
\begin{align}
W_-(z) y(z/q)+W_+(z)y(qz)-R(z)y(z)=0\,,
\label{eq:Lax_equation_of_Garnier}
\end{align}
where 
\begin{align}
&W_-(z)\equiv A(k/z)B(z)F(qz)[k/q^2z^2],\notag\\
&W_+(z)\equiv A(qz)B(k/z)F(z)[k/z^2],
\end{align}
with\footnote{The $``N"$ in $A(z)$, $B(z)$ and $F(z)$ defined in \cite{MR4120359} has been shifted to $``N+3"$ in according with the $\sprm(N)$ curve.}
\begin{align}
A(z)\equiv \prod_{j=1}^{N+4}[z/a_j]\,,\quad\quad
B(z)\equiv \prod_{j=1}^{N+4}[z/b_j]\,,\quad\quad
F(z)\equiv Cz\prod_{j=1}^{N+1}[z/\lambda_j][k/z\lambda_j].
\end{align}
We further follow the parametrizations in \cite{MR4120359} that
\begin{align}
a_i=q M_i\,,\quad\quad b_i=q M_{N + 4 + i}\,,\quad\quad k=pq^2\,,\quad {\rm and}\quad \lambda_i= q\nu_i\,,
\end{align}
to rewrite the above equation in terms of $R_p(X)$. Notice that 
\begin{align}
&W_-(z)=\prod_{j=1}^{N+4}R_p(XM_{j}q^{-1/2})R_p(XM_{N+1+j}^{-1}q^{-1/2})\notag\\
&\qquad\qquad\times CQ^{1/2} q^{3/2}XR_p(X^2 q Q^{1/2})\prod_{j=1}^{N+1}R_p(X\nu_i^{\pm 1}q^{1/2}),\notag\\
&W_+(z)=\prod_{j=1}^{N+4}R_p(XM_{j}^{-1}q^{1/2})R_p(XM_{N+1+j}q^{1/2})\notag\\
&\qquad\qquad\times CQ^{1/2}q^{1/2}XR_p(X^2 q^{-1}Q^{1/2})\prod_{j=1}^{N+1}R_p(X\nu_i^{\pm 1} q^{-1/2})\,.
\end{align}
Now define an additional function
\begin{align}
D(z)&\equiv C p^{-1}q^{-1}z^3[k/z^2][k/qz^2][k/q^2z^2]\prod_{j=1}^{N+1}[z/\lambda_j][k/qz\lambda_j]\notag\\
&=C p^{1/2}q^{1/2}X^3R_p(X^2 q^{1/2})R_p(X^2 p^{\pm 1} q^{1/2})\prod_{j=1}^{N+1}R_p(X\nu_i q^{1/2})R_p(X\nu_i^{-1} q^{-1/2})\,.
\end{align}
Therefore we have
\begin{align}
\frac{W_-(z)}{D(z)}&=q X^{-2}\prod_{j=1}^{N+1}\frac{R_p(X\nu_i^{-1}q^{1/2})}{R_p(X\nu_i^{-1}q^{-1/2})}\times\frac{\prod_{j=1}^{N+4}R_p(XM_{j}q^{-1/2})R_p(XM_{N+4+j}^{-1}q^{-1/2})}{R_p(X^2p^{1/2})R_p(X^2q^{-1}p^{1/2})},\notag\\\notag\\
\frac{W_+(z)}{D(z)}&=X^{-2}\prod_{j=1}^{N+1}\frac{R_p(X\nu_i q^{-1/2})}{R_p(X\nu_i q^{1/2})}\times\frac{\prod_{j=1}^{N+4}R_p(XM_{j}^{-1}q^{1/2})R_p(XM_{N+4+j}q^{1/2})}{R_p(X^2p^{1/2})R_p(X^2qp^{1/2})}\,.
\end{align}
Now we assign $\{\nu_j\}$ equal to the last $N+1$ $\{M_{j}\}$, i.e.
\begin{align}
\nu_j=M_{N+7+j}\,, \quad {\rm for} \quad j=1,\dots, N+1
\label{vM}
\end{align}
to cancel $R_p(X\nu_i^{-1}q^{-1/2})$ with $R_p(XM_{N+ 7 + i}^{-1}\,q^{-1/2})$ in $\frac{W_-(z)}{D(z)}$,
and $R_p(X\nu_i q^{1/2})$ with $R_p(XM_{N + 7 + i}\,q^{1/2})$ in $\frac{W_+(z)}{D(z)}$, and further shift 
\begin{align}
M_{N+7+j}\longrightarrow q M_{N+7+j}\,, \quad {\rm for} \quad j=1,\dots, N + 1\,.
\label{vM_shift}
\end{align}
Therefore we have
\begin{align}
\frac{W_-(z)}{D(z)}&=q X^{-2}\frac{\prod_{j=1}^{N+4}R_p(XM_{j}q^{-1/2})R_p(XM_{N+4+j}^{-1}q^{-1/2})}{R_p(X^2p^{1/2})R_p(X^2q^{-1}p^{1/2})}\equiv q X^{-2}\widetilde V(x),\notag\\\notag\\
\frac{W_+(z)}{D(z)}&=X^{-2}\frac{\prod_{j=1}^{N+4}R_p(XM_{j}^{-1}q^{1/2})R_p(XM_{N+4+j}q^{1/2})}{R_p(X^2p^{1/2})R_p(X^2qp^{1/2})}\notag\\
&=qX^{2}\frac{\prod_{j=1}^{N+4}R_p(X^{-1}M_{j}q^{-1/2})R_p(X^{-1}M_{N+4+j}^{-1}q^{-1/2})}{R_p(X^{-2}p^{1/2})R_p(X^{-2}q^{-1}p^{1/2})}=q X^2\widetilde V(-x)\,.
\end{align}
where we have applied the even and quasi $\tau$-periodic properties of $R_p(x)$, i.e.
\begin{align}
R_p(X^{-1})=R_p(X)\,, \quad {\rm and} \quad R_p(X p^{-1/2})=-X R_p(X p^{1/2})\,.
\end{align}
With a similar transformation noticed in \cite{MR4120359},
\begin{align}
g(x)=R_q(Xq^{-1/2})R_q(Xq^{1/2})\prod_{j=1}^{N+4}\frac{\Gamma_{p, q}(X M_{j}q^{1/2}p^{1/2})}{\Gamma_{p, q}(X M_{j}^{-1}q^{1/2}p^{1/2})}\,,
\label{eq:gt_1}
\end{align}
one can show that
\begin{align}
g(x)^{-1} Y \cdot g(x)&=q^{-1}X^2\prod_{j=1}^{N+4}\frac{[X M_{j}^{-1}q^{-1/2}p^{1/2}]}{[X M_{j}q^{-1/2}p^{1/2}]}\cdot Y=q^{-1}X^2\prod_{j=1}^{N+4}\frac{R_p(X M_{j}^{-1}q^{-1/2})}{R_p(X M_{j} q^{-1/2})}\cdot Y\,,\notag\\\notag\\
g(x)^{-1}\cdot Y^{-1} \cdot g(x)&=q^{-1}X^{-2}\prod_{j=1}^{N+4}\frac{[X M_{j}q^{1/2}p^{1/2}]}{[X M_{j}^{-1}q^{1/2}p^{1/2}]}\cdot Y^{-1}=q^{-1}X^{-2}\prod_{j=1}^{N+4}\frac{R_p(X M_{j}q^{1/2})}{R_p(X M_{j}^{-1}q^{1/2})}\cdot Y^{-1}\,,
\end{align}
where the $q$-shift operator $Y$ is defined as
\begin{align}
Y:\quad X\longrightarrow q^{-1}X\,, \quad {\rm or} \quad x\longrightarrow x-\epsilon_1\,,
\end{align}
and we also used the property of the elliptic Gamma function
\begin{align}
\Gamma_{p, q}(q X)=[X]\,\Gamma_{p, q}(X)\,, \quad {\rm or} \quad \Gamma_{p, q}(q^{-1}X)=\frac{\Gamma_{p, q}(X)}{[q^{-1}X]}\,.
\end{align}
Therefore the shifted part of the Lax equation,
\begin{align}
\mathcal D_{\rm shift}\equiv\frac{W_-(z)}{D(z)}\cdot Y+\frac{W_+(z)}{D(z)}\cdot Y^{-1}\,,
\end{align}
 in the Garnier models can be recast via the gauge transformation $g(x)$ as
 \begin{align}
 \mathcal D_{\rm shift} \longrightarrow\,g(x)^{-1}\cdot\mathcal D_{\rm shift}\cdot g(x)=V_{\rm NRY}(x)\cdot Y+V_{\rm NRY}(-x)\cdot Y^{-1}\,,
 \end{align}
with
\begin{align}
V_{\rm NRY}(x)\equiv\frac{\prod_{j=1}^{2N+8}R_p(XM_{j}^{-1}q^{-1/2})}{R_p(X^2p^{1/2})R_p(X^2q^{-1}p^{1/2})}\,,
\end{align}
where $V_{\rm NRY}(x)$ generalizes the result in \cite{MR4120359}.

\paragraph{Additive part}
Now we turn to the additive part of the difference operator in the Garnier models,
\begin{align}
\mathcal D_{\rm add}\equiv -\frac{R(z)}{D(z)}\,.
\end{align}
Since this piece is only a function, the previous gauge transformation $g(x)$ keep it intact. Notice that
\begin{align}
R(z)=S_1(z)+S_2(z)+S_3(z)\,,
\end{align}
where
\begin{align}
&S_1(z)\equiv U(z)F(qz)G(k/z)[k/q^2z^2]/G(z)\,,\notag\\
&S_2(z)\equiv U(k/qz)F(z)G(qz)[k/z^2]/G(k/qz)\,,\notag\\
&S_3(z)\equiv -F(z)F(qz)\overline F(z)[k/z^2][k/q z^2][k/q^2z^2]/G(z)G(k/qz)\,,
\end{align}
with
\begin{align}
G(z)\equiv z\prod_{j=1}^{N+2}[z/\xi_j]\,,\quad
{\rm and}\quad \prod_j^{N+2}\xi_j=l\,, \quad 
k^2l^2=q\prod_{j=1}^{N+4}a_j b_j\,.
\label{constraint}
\end{align}
With these preparations, we rewrite $S_i(z)$ in terms of function $R_p(x)$. First for $S_1(z)$, we have
\begin{align}
&U(z)=A(z)B(z)=\prod_{j=1}^{2N+8}R_p(XM_j^{-1}q^{-1/2})\,,\notag\\
&F(qz)=Cqz\prod_{j=1}^{N+1}[qz/\lambda_j][pq/z\lambda_j]=Cp^{1/2}q^{3/2}X\prod_{j=1}^{N+1}R_p(X\nu_j^{\pm 1}q^{1/2})\,,\notag\\
&G(z)=z\prod_{j=1}^{N+2}[z/\xi_i]=p^{1/2}q^{1/2}X\prod_{j=1}^{N+2}R_p(X\alpha_j^{-1}q^{-1/2})\,,\notag\\
&G(k/z)=k/z\prod_{j=1}^{N+21}[k/z\xi_i]=p^{1/2}q^{3/2}X^{-1}\prod_{j=1}^{N+2}R_p(X\alpha_j q^{-1/2})\,,
\end{align}
where we have defined
\begin{align}
\xi_j=q\alpha_j\,.
\end{align}
Therefore
\begin{align}
S_1(z)&=Cp^{1/2}q^{5/2}X^{-1}R_p(X^2qp^{1/2})\prod_{j=1}^{2N+8}R_p(XM_j^{-1}q^{-1/2})\notag\\
&\quad\times\prod_{j=1}^{N+1}R_p(X\nu_j^{\pm 1}q^{1/2})\times\prod_{j=1}^{N+2}\frac{R_p(X\alpha_j q^{-1/2})}{R_p(X\alpha_j^{-1}q^{-1/2})}\,,
\end{align}
and
\begin{align}
E(x)&\equiv\frac{S_1(z)}{D(z)}=q^2X^{-4}
\prod_{j=1}^{N-2}\frac{R_p(X\nu_j^{- 1}q^{1/2})}{R_p(X\nu_j^{- 1}q^{-1/2})}\times\frac{\prod_{j=1}^{2N+2}R_p(XM_j^{-1}q^{-1/2})}{R_p(X^2p^{1/2})R_p(X^2q^{-1}p^{1/2})}\notag\\
&\quad\qquad\qquad\times\prod_{j=1}^{N-1}\frac{R_p(X\alpha_j q^{-1/2})}{R_p(X\alpha_j^{-1}q^{-1/2})}\,.
\end{align}
Similarly for $S_2(z)$, we have
\begin{align}
&U(k/qz)=\prod_{j=1}^{2N+8}R_p(XM_jq^{1/2})\,,\notag\\
&F(z)=Cp^{1/2}q^{1/2}X\prod_{j=1}^{N+1}R_p(X\nu_j^{\pm 1}q^{-1/2})\,,\notag\\
&G(qz)=p^{1/2}q^{3/2}X\prod_{j=1}^{N+2}R_p(X\alpha_j^{-1}q^{1/2})\,,\notag\\
&G(k/qz)=p^{1/2}q^{1/2}X^{-1}\prod_{j=1}^{N+2}R_p(X\alpha_j q^{1/2})\,.
\end{align}
Therefore
\begin{align}
S_2(z)=Cp^{1/2}q^{3/2}X^3R_p(X^2q^{-1}p^{1/2})
\prod_{j=1}^{2N+8}R_p(XM_jq^{1/2})
\prod_{j=1}^{N+1}R_p(X\nu_j^{\pm 1}q^{-1/2})
\prod_{j=1}^{N+2}\frac{R_p(X\alpha_j^{-1}q^{1/2})}{R_p(X\alpha_j q^{1/2})}\,,
\end{align}
and
\begin{align}
\frac{S_2(z)}{D(z)}&=q\prod_{j=1}^{N+1}\frac{R_p(X\nu_j q^{-1/2})}{R_p(X\nu_j q^{1/2})}\times\frac{\prod_{j=1}^{2N+8}R_p(XM_j q^{1/2})}{R_p(X^2p^{1/2})R_p(X^2 q p^{1/2})}
\times\prod_{j=1}^{N+2}\frac{R_p(X\alpha_j^{-1} q^{1/2})}{R_p(X\alpha_j q^{1/2})}\notag\\\notag\\
&=q^2X^4
\prod_{j=1}^{N+1}\frac{R_p(X^{-1}\nu_j^{-1} q^{1/2})}{R_p(X^{-1}\nu_j^{-1} q^{-1/2})}
\times
\frac{\prod_{j=1}^{2N+8}R_p(X^{-1}M_j^{-1} q^{-1/2})}{R_p(X^{-2}p^{1/2})R_p(X^{-2} q^{-1} p^{1/2})}
\times
\prod_{j=1}^{N+2}\frac{R_p(X^{-1}\alpha_j q^{-1/2})}{R_p(X^{-1}\alpha_j^{-1} q^{-1/2})}\notag\\\notag\\
&=E(-x)\,.
\end{align}
At last, we spell out the third piece of $R(z)/D(z)$,
\begin{align}
V_{\rm e}(x)\equiv\frac{S_3(z)}{D(z)}=C\overline Cq\frac{\prod_{j=1}^{N+1}R_p(X^{\pm 1}M_{N+7+j}q^{-1/2})R_p(X^{\pm 1}\overline M_{N+7+j}q^{-1/2})}{\prod_{j=1}^{N+2}R_p(X^{\pm 1}\alpha_jq^{1/2})}\,.
\end{align}

Now we turn to discuss the poles in $E(x)+E(-x)+V_{\rm e}(x)$. Recall \eqref{vM} and \eqref{vM_shift}, we simplify $E(x)$ as
\begin{align}
E(x)=q^2X^{-4}
\frac{\prod_{j=1}^{2N+8}R_p(XM_j^{-1}q^{-1/2})}{R_p(X^2p^{1/2})R_p(X^2q^{-1}p^{1/2})}
\times\prod_{j=1}^{N+2}\frac{R_p(X\alpha_j q^{-1/2})}{R_p(X\alpha_j^{-1}q^{-1/2})}\,.
\end{align}
We are aiming to show $E(x)$ is a theta function of degree $2N$. Notice that
\begin{align}
\frac{E(Xp^{-1/2})}{E(Xp^{1/2})}=p^{4}\frac{X^{2N+8}q^{-N-4}\prod_{j=1}^{2N+8}M_j^{-1}}{X^8 p^2 q^{-2}}\times\prod_{j}^{N+2}\alpha_j^2 =
p^2q^{-N-2}X^{2N}\frac{\prod_{j=1}^{N+2}\alpha_j^2}{\prod_{j=1}^{2N+8}M_j}\,.
\end{align}
Further using the constraint \eqref{constraint}, we have
\begin{align}
\prod_{j=1}^{N+2}\alpha_j^2=q^{-2(N+2)}l^2=p^{-2}q^{-4}q^{-2(N+2)}q^{2N+9}q^{N+1}\prod_{j=1}^{2N+8}M_j=p^{-2}q^{N+2}\prod_{j=1}^{2N+8}M_j\,,
\end{align}
where the additional $q^{N+1}$ in the second equality is due to the shift of $N+1$ number of $M_j$, see \eqref{vM_shift}. Therefore we have
\begin{align}
\frac{E(Xp^{-1/2})}{E(Xp^{1/2})}=X^{2N}\,,
\label{eq:quasi-perodicity_of_E}
\end{align}
Now we compute the residues of $E(x)$ at poles at
\begin{align}
x_a=\frac{\epsilon_1}{2}-u_a\,,
\end{align}
where $u_a=\{0,\, \frac{1}{2},\, \frac{\tau+1}{2},\, \frac{\tau}{2}\}$.
\begin{align}
&\rho_1\equiv{\rm Res}\, E(x)\vert_{x=x_1}=\frac{\prod_{j=1}^{2N+8}R_p(M_j)}{4\pi i\kappa^2 R_p(qp^{1/2})}
=\frac{\prod_{j=1}^{2N+8}\vartheta_4(\mu_j)}{4\pi i\kappa^{2N+9}\vartheta_4(\epsilon_1+\frac{\tau}{2})}\,,\notag\\\notag\\
&\rho_2\equiv{\rm Res}\, E(x)\vert_{x=x_2}=\frac{\prod_{j=1}^{2N+8}R_p(-M_j)}{4\pi i\kappa^2 R_p(qp^{1/2})}
=\frac{\prod_{j=1}^{2N+8}\vartheta_3(\mu_j)}{4\pi i\kappa^{2N+9}\vartheta_4(\epsilon_1+\frac{\tau}{2})}\,,\notag\\\notag\\
&\rho_3\equiv{\rm Res}\, E(x)\vert_{x=x_3}=-\frac{p^2\prod_{j=1}^{2N+8}R_p(-M_jp^{1/2})}{4\pi i \kappa^{2}R_p(qp^{-1/2})}\prod_{j=1}^{N+2}\frac{R_p(-\alpha_jp^{-1/2})}{R_p(-\alpha_jp^{1/2})}
=\frac{p^{-N/4}q^{N/2}\prod_{j=1}^{2N+8}\vartheta_2(\mu_j)}{4\pi i\kappa^{2N+9}\vartheta_4(\epsilon_1+\tau/2)}\,,\notag\\\notag\\
&\rho_4\equiv{\rm Res}\, E(x)\vert_{x=x_4}
=-\frac{p^2\prod_{j=1}^{2N+8}R_p(M_jp^{1/2})}{4\pi i \kappa^{2}R_p(qp^{-1/2})}\prod_{j=1}^{N+2}\frac{R_p(\alpha_jp^{-1/2})}{R_p(\alpha_jp^{1/2})}
=\frac{p^{-N/4}q^{N/2}\prod_{j=1}^{2N+8}\vartheta_1(\mu_j)}{4\pi i\kappa^{2N+9}\vartheta_4(\epsilon_1+\tau/2)}\,,
\label{eq:residue_of_additive}
\end{align}
where $\kappa\equiv\prod_{n=1}^\infty(1-p^n)$. Notice that the additional factor $p^{-N/4}q^{N/2}$ is due to the quasi-perodicity \eqref{eq:quasi-perodicity_of_E}. Similarly, for the residues of $E(-x)$ at poles of
\begin{align}
x=-x_a=-\frac{\epsilon_1}{2}+u_a\,,
\end{align}
we have
\begin{align}
{\rm Res}\, E(-x)\vert_{x=-x_a}=-\rho_a\,.
\end{align}
It's also worth mentioning that, for the poles in $R_p(X^2p^{\pm 1/2})$ in $E(\pm x)$, the residues are canceled due to the evenness of $E(x)+E(-x)$. There are also no poles in $V_{\rm e}(x)$ as argued in \cite{MR4120359}.

\subsection{$\sprm(N)$ quantum curves}
Now we turn to discuss the $\sprm(N)$ quantum curves. We aim to show the equivalence of the $\sprm(N)$ quantum curves and the Lax equation of the Garnier system \eqref{eq:Lax_equation_of_Garnier} discussed in previous section.

\paragraph{Additive part}
We aim to identify the additive part in the $\sprm(N)$ quantum curves by examining the residues of the singular part in the $\sprm(N)$ quantum curves and comparing them with the residues of the additive part in the elliptic Garnier system. Recall that the analogous additive part in the $\sprm(N)$ quantum curves is given by the singular part $\mathcal{V}(x)$ of the whole codimension four defect partition function, which is a singular part of the product of the perturbative contribution and the one-instanton correction to the normalized Wilson surface defect. As a result, the residues of the singular part are contributed by the residues of
\begin{align}
\widetilde\chi(x)=W_{\rm pert}\times \mathfrak{q}\,\chi_1\,,
\end{align}
where from eq.\,(\ref{eq:pert_Wilson}) and (\ref{eq:1-inst_chi})
\begin{align}
&W_{\rm pert}=\mathfrak{q}^{-1}\prod_{i=1}^N\frac{\vartheta_1(x\pm\alpha_i)}{i\eta}\,,\notag\\
&\chi_{1}=-\frac{1}{2\eta^{6}\vartheta_1(\epsilon_{1})\vartheta_1^\prime(0)}\sum_a\frac{\prod_{f=1}^{2N+8}\vartheta_a(m_f)}{\prod_{i=1}^N\vartheta_a\left(\pm\alpha_i+\frac{\epsilon_1}{2}\right)}\vartheta_a^\Delta(x)\,.
\end{align}
The product of the two parts is independent of the tensor fugacity $\mathfrak{q}$; thus, we will omit the tensor fugacity in the following calculations, as long as doing so does not cause any confusion.
For further comparison to the Lax equation in the Garnier systems, we assign $m_f$ as
\begin{align}
m_f=\mu_f+\frac{\tau}{2}\,.
\label{eq:mass_rep}
\end{align}
With this parametrization, we have
\begin{align}
\chi_{1}&=-\frac{(-1)^N p^{-(N+4)/4}\prod_{f=1}^{2N+8}M_f^{-1/2}}{2\eta^6\vartheta_1^\prime(0)\vartheta_1(\epsilon_1)}\sum_{a=1}^4\frac{\prod_{f=1}^{2N+8}\vartheta_{\sigma(a)}(\mu_f)}{\prod_{i=1}^N\vartheta_{a}(\pm\alpha_i+\epsilon_1/2)}\cdot\vartheta_a^{\Delta}(x)\notag\\\notag\\
&=\frac{(-1)^N p^{-(N+6)/4}q^{-1/2}\prod_{f=1}^{2N+8}M_f^{-1/2}}{4\pi i\kappa^9\vartheta_4(\epsilon_1+\tau/2)}\sum_{a=1}^4\frac{\prod_{f=1}^{2N+8}\vartheta_{\sigma(a)}(\mu_f)}{\prod_{i=1}^N\vartheta_{a}(\pm\alpha_i+\epsilon_1/2)}\cdot\vartheta_a^{\Delta}(x)
\end{align}
where $\sigma=(14)(23)$ permuting the set $\{1,\, 2,\, 3,\, 4\}$. In addition, recall the perturbative piece
\begin{align}
W_{\rm pert}=\prod_{i=1}^N\frac{\vartheta_1(x\pm\alpha_i)}{i\eta}=\frac{(-1)^N p^{-N/12}}{\kappa^{2N}}\prod_{i=1}^N\vartheta_1(x\pm\alpha_i)\,.
\end{align}
Overall, we have
\begin{align}
\widetilde\chi(x)=\mathcal C\times\frac{\prod_{i=1}^N\vartheta_1(x\pm\alpha_i)}{4\pi i\kappa^{2N+9}\vartheta_4(\epsilon_1+\tau/2)}\sum_{a=1}^4\frac{\prod_{f=1}^{2N+8}\vartheta_{\sigma(a)}(\mu_f)}{\prod_{i=1}^N\vartheta_{a}(\pm\alpha_i+\epsilon_1/2)}\cdot\vartheta_a^{\Delta}(x)\,,
\end{align}
with the constant
\begin{align}
\mathcal C= p^{-(2N+9)/6}q^{-1/2}\prod_{f=1}^{2N+8}M_f^{-1/2}\,.
\end{align}
Let us study the poles of $\chi(x)$ and the corresponding residues. Notice that all poles in $\chi(x)$ are from the term $\vartheta_I^{\Delta}(x)$, which are
\begin{align}
x_a=\frac{\epsilon_1}{2}-u_a\,,
\end{align}
in the toric lattice. The residue of $\vartheta_a^{\Delta}(x)$ is precisely
\begin{align}
{\rm Res}\, \vartheta_a^\Delta(x)\vert_{x=\pm x_b}=\pm \delta_{ab}\,.
\label{eq:residue_of_theta_Delta}
\end{align}
Furthermore, the perturbative contribution $W_{\rm pert}$ is given by $\prod_{i=1}^N\vartheta_1(x\pm \alpha_i)$ in $\chi(x)$ 
\begin{align}
&\prod_{i=1}^N\vartheta_1(x\pm \alpha_i)\Big\vert_{x=\pm x_1}=\prod_{i=1}^N\vartheta_1(\pm \alpha_i+\epsilon_1/2)\,,\notag\\
&\prod_{i=1}^N\vartheta_1(x\pm \alpha_i)\Big\vert_{x=\pm x_2}=\prod_{i=1}^N\vartheta_2(\pm\alpha_i+\epsilon_1/2)\,,\notag\\
&\prod_{i=1}^N\vartheta_1(x\pm \alpha_i)\Big\vert_{x=\pm x_3}=p^{-N/4}q^{N/2}\prod_{i=1}^N\vartheta_3(\pm\alpha_i+\epsilon_1/2)\,,\notag\\
&\prod_{i=1}^N\vartheta_1(x\pm \alpha_i)\Big\vert_{x=\pm x_4}=(-1)^N p^{-N/4}q^{N/2}\prod_{i=1}^N\vartheta_4(\pm\alpha_i+\epsilon_1/2)\,,
\label{eq:residue_of_Wpert}
\end{align}
where there is an unwanted factor $(-1)^N$ in the last equation. It can be absorbed by shifting $N$ mass parameter $\mu_f\rightarrow -\mu_f$ and the odd property of $\vartheta_1$ in the instanton piece. With this redefinition and applying \eqref{eq:residue_of_theta_Delta} and \eqref{eq:residue_of_Wpert}, one can honestly find, beside the prefactor $\mathcal C$, that
\begin{align}
\mathcal C^{-1}\times {\rm Res}\,\widetilde\chi(x)\Big\vert_{x=\pm x_a}= \pm\rho_a\,,
\end{align}
where $\rho_a$ are given by \eqref{eq:residue_of_additive}. Moreover the quasi-elliptic property of $\chi(x)$ is precisely
\begin{align}
\frac{\widetilde\chi(x-\tau/2)}{\widetilde\chi(x+\tau/2)}=X^{2N}\,.
\end{align}
Therefore we claim that $\mathcal C^{-1}\widetilde\chi(x)$ and $E(x)+E(-x)+V_{\rm e}(x)$ have same poles with same residues.

\paragraph{Shift part} Now we come to the shift part of the $\sprm(N)$ quantum curve. Recall that the shift part is given by eq.\eqref{eq:SpN_curve} with
\begin{align}
V(x)=\prod_{f=1}^{2N+8}\frac{\theta_1(x-m_f-\epsilon_1/2)}{\theta_1(2x)\theta_1(2x-\epsilon_1)}=-\frac{(-1)^N}{\eta^{2N+6}}\prod_{f=1}^{2N+8}\frac{\vartheta_1(x-m_f-\epsilon_1/2)}{\vartheta_1(2x)\vartheta_1(2x-\epsilon_1)}\,.
\end{align}
Applying \eqref{eq:mass_rep} to $V(x)$, one can recast it in terms of $R_p(x)$ as
\begin{align}
V(x)=\mathcal C\times q^{-(N+1)/2}X^{N+2}\frac{\prod_{j=1}^{2N+2}R_p(XM_{j}^{-1}q^{-1/2})}{R_p(X^2p^{1/2})R_p(X^2q^{-1}p^{1/2})}=\mathcal C\times q^{-(N+2)/2}X^{N+2}\cdot V_{\rm NRY}(x)\,.
\end{align}
Similarly, for $V(-x)$, we have
\begin{align}
V(-x)=\mathcal C\times q^{-(N+2)/2}X^{-(N+2)}\cdot V_{\rm NRY}(-x)\,.
\end{align}
Notice in \eqref{eq:gt_1}, the gauge transformation
\begin{align}
f(x)=R_q(Xq^{-1/2})R_q(Xq^{1/2})
\end{align}
satisfies
\begin{align}
&f(x)^{-1}\cdot Y \cdot f(x)=q^{-1}X^2\cdot Y\,,\notag\\
&f(x)^{-1}\cdot Y^{-1} \cdot f(x)=q^{-1}X^{-2}\cdot Y^{-1}\,.
\end{align}
We thus introduce a gauge transformation
\begin{align}
h(x)=f(x)^{-(N+2)/2}\,,
\end{align}
with the transformation property
\begin{align}
&h(x)^{-1}\cdot Y \cdot h(x)=q^{(N+2)/2}X^{-(N+2)}\cdot Y\,,\notag\\
&h(x)^{-1}\cdot Y^{-1} \cdot h(x)=q^{(N+2)/2}X^{N+2}\cdot Y^{-1}\,.
\end{align}
Therefore, applying the above gauge transformation, and dividing the prefactor $\mathcal C$ in the $\sprm(N)$ quantum curves \eqref{eq:SpN_curve}, we succeed in showing that the $\sprm(N)$ quantum curves coincide with the Lax equation \eqref{eq:Lax_equation_of_Garnier} of the elliptic Garnier systems.
\section{RG Flows to 5d}\label{sec:RG}
The circle compactification of 6d $D$-type minimal conformal matter is effectively described by 5d KK theory with gauge group $\spn$ or $\surm(N+1)$ plus $(2N+6)$ fundamental flavors \cite{Hayashi:2015fsa,Hayashi:2015zka,Jefferson:2018irk,Bhardwaj:2020gyu}. We can integrate out the masses of the fundamental flavors to get theories with a lower number of fundamental flavors. In this section, we focus on the flows to 5d $\sprm(N)$ gauge theories. We take the limits of the 6d $\sprm(N-1)$ curves and obtain the curves for 5d $\spn$ theories with $N_f<2N+6$ fundamental flavors.
\subsection{General cases}
We begin with the KK theory $\spn+(2N+6)\mathsf{F}$, where the corresponding 6d theory is $\sprm(N-1)+(2N+6)\mathsf{F}$. As discussed in \cite{Kim:2014dza,Hayashi:2016abm,Yun:2016yzw}, we need to first turn on the holonomy of the Wilson line along the compactified circle, which shift one of the mass parameter and the tensor parameter $\phi_0$ by
\begin{align}\label{eq:map1}
    m_{1}\rightarrow m_{1}+2\pi i \tau,\quad \phi_0 \rightarrow \phi_0 +i\pi \tau +m_{1}\,.
\end{align}
Subsequently, the maps between Coulomb parameters can be deduced from the group decomposition 
\begin{align}
    \sprm(N) \rightarrow \mathrm{SU}(2)\times \sprm(N-1),
\end{align}
such that the $\sprm(N-1)$ describes the gauge group of the 6d theory on the tensor branch, so we have the maps for the Coulomb parameters in the basis spanned by the fundamental weights
\begin{align}
    \phi_0\rightarrow \phi_1^{\text{5d}},\quad\quad \phi_j\rightarrow \phi_{j+1}^{\text{5d}}-\phi_1^{\text{5d}},\quad j=1,\cdots N-1,
\end{align}
and the $(2N+6)$ mass parameters become the mass parameters $m_l,l=1,\cdots, 2N+6$ in the 5d description and the complex structure parameter $2\pi i \tau$ becomes the 5d instanton counting parameter $m_0$. Further flow from $\sprm(N)$ with $N_f$ flavors to $\sprm(N)$ with $N_f-1$ flavors is amount to take the limit in the curve
\begin{align}\label{eq:map2}
    m_0\rightarrow m_0+m_{N_f},\quad\quad m_{N_f}\rightarrow \infty.
\end{align}
With all these maps in mind, we are now ready to write down the curves for the 5d KK theories.
\begin{align} 
\left(V(x) Y+V(-x)  Y^{-1} + \chi(x)\right) \Psi(x)=0\,,
\label{eq:SpN_curvefff2}
\end{align}
which is, and should be, the same as the 6d curve \eqref{eq:SpN_curvefff2}. Now the 6d codimension-4 defect partition function can also be realized as the combination of the Wilson loop expectation values of 5d theory, where the representations of the Wilson loops can be read from the leading term expansion in terms of Coulomb parameters. 

\paragraph{\underline{$\sprm(N)+(2N+5)\mathsf{F}$}}
In this subsection, we derive the curve for 5d $\sprm(N)$ theory with $(2N+5)$ fundamental flavors. Note that by using the Weyl symmetry of the affined $D$-type flavor group,
\begin{align}
    m_1\rightarrow m_1 -2\pi i \tau, \quad m_{2N+6}\rightarrow m_{2N+6} -2\pi i \tau,
\end{align}
the maps \eqref{eq:map1} and \eqref{eq:map2} can be alternatively written as
\begin{align}
    &m_{2N+6}\rightarrow -m_0,\quad\quad m_i \rightarrow m_i,\quad i=1,\cdots 2N+5,\nonumber\\
    &\phi_0\rightarrow \phi_0 -\frac{1}{2}m_0,\quad\quad\tau\rightarrow \infty.
\end{align}
We then derive the curve
\begin{align}\label{eq:curve5}
    Y+\frac{\prod_{l=0}^{2N+5}\sh(x\pm m_l+\epsilon_1/2)}{\sh(2x+\epsilon_1\pm \epsilon_1)\sh(2x+\epsilon_1)^2}Y^{-1}+V_0(x) = (-1)^{N-1}e^{-\frac{1}{2}m_0}\sum_{j=1}^{N}(-1)^{j+1}\ch(2x)^{N-j} H_j,
\end{align}
with $V_0(x)$ as a potential from the Coulomb independent part of one-instanton codimension-four defect partition function,
\begin{align}\label{eq:V05}
    V_0(x)=&-(-1)^N\frac{\prod_{l=0}^{2N+5}\ch(m_l)}{2\,\ch(x\pm \epsilon_1/2)}+\frac{\prod_{l=0}^{2N+5}\sh(m_l)}{2\,\sh(x\pm \epsilon_1/2)}-\ch(2x)^{N}\sum_{l=0}^{2N+5}\ch(2m_l)\nonumber\\
    &\quad\quad\quad\quad+\ch(\epsilon_1)\,\ch(2x)^{N-1}\left(\ch(4x)+(N-1)\,\sh(\epsilon_1)^2\right)\,,
\end{align}
where $\sh(x)=e^{\frac{x}{2}}-e^{-\frac{x}{2}},\ch(x)=e^{\frac{x}{2}}+e^{-\frac{x}{2}}$ and 
\begin{align}
    \chi_{j}(x)=\frac{e^{(j+1)x}-e^{-(j+1)x}}{e^x-e^{-x}},
\end{align}
is the character of $SU(2)$ with the highest weight $j$. Notice that $H_j$ is the eigenvalue of the Hamiltonian of the corresponding integrable system, which should be the NS limit of the Wilson loop expectation value in the {\it{orbit}} of the $j$-th fundamental weight of the gauge group $\sprm(N)$. In the study of topological string theory, the quantum curve coincides with the quantum curve in the topological string B-model and the parameters $H_i$ we defined here are the most natural complex structure parameters in the B-model. We will have a discussion for more details on this in Section \ref{sec:TSTandWilson}.

One important property of the quantum curve \eqref{eq:curve5} is that if we absorb the factor $e^{-\frac{1}{2}m_0}$ into the Hamiltonians, the curve has the manifest $SO(4N+12)$ global symmetry, which is the enhanced global symmetry of $\sprm(N)+(2N+5)\mathsf{F}$ theory.
\paragraph{\underline{$\sprm(N)+(2N+4)\mathsf{F}$}}
According to the map \eqref{eq:map2}, we now take shift 
\begin{align}\label{eq:limit5to4}
    m_0\rightarrow m_0+m_{2N+5}, 
\end{align}
then the leading terms under the limit $m_{2N+5}\rightarrow \infty$, give the quantum curve of $\sprm(N)+(2N+4)\mathsf{F}$
\begin{align}\label{eq:curve4}
    Y+\frac{\prod_{l=1}^{2N+4}\sh(x\pm m_l+\epsilon_1/2)}{\sh(2x+\epsilon_1\pm \epsilon_1)\sh(2x+\epsilon_1)^2}Y^{-1}+V_0^{\prime}(x) =(-1)^{N-1} e^{-\frac{1}{2}m_0}\sum_{j=1}^{N}(-1)^{j+1}\ch(2x)^{N-j} H_j,
\end{align}
where
\begin{align}
    V_0^{\prime}(x)=&-(-1)^N\frac{\prod_{l=1}^{2N+4}\ch(m_l)}{2\,\ch(x\pm \epsilon_1/2)}+\frac{\prod_{l=1}^{2N+4}\sh(m_l)}{2\,\sh(x\pm \epsilon_1/2)}-\ch(2x)^{N}\ch(m_0).
\end{align}
If we absorb the factor $e^{-\frac{1}{2}m_0}$ into the Hamiltonians, the curve \eqref{eq:curve4} has the manifest $SO(4N+8)\times SU(2)$ global symmetry, which is the enhanced global symmetry of $\sprm(N)+(2N+4)\mathsf{F}$ theory.
\paragraph{\underline{$\sprm(N)+N_f\mathsf{F}$, $N_f\leq 2N+3$}}
We can further take the massive limit of mass parameters the curve for $\sprm(N)+N_f\mathsf{F}$ with $N_f\leq 2N+3$ can be universally written as
\begin{align}\label{eq:curve3}
    Y+\frac{(-1)^{N_f}\prod_{l=1}^{N_f}\sh(x\pm m_l+\epsilon_1/2)}{\sh(2x+\epsilon_1\pm \epsilon_1)\sh(2x+\epsilon_1)^2}Y^{-1}+V_0^{\prime\prime}(x) = (-1)^{N-1}e^{-\frac{1}{2}m_0}\sum_{j=1}^{N}(-1)^{j+1}\ch(2x)^{N-j} H_j,
\end{align}
where
\begin{align}
    V_0^{\prime\prime}(x)=&-(-1)^N\frac{\prod_{l=1}^{N_f}\ch(m_l)}{2\,\ch(x\pm \epsilon_1/2)}+(-1)^{N_f}\frac{\prod_{l=1}^{N_f}\sh(m_l)}{2\,\sh(x\pm \epsilon_1/2)}-e^{-\frac{1}{2}m_0}\ch(2x)^{N}.
\end{align}
In particular, when $N_f=0$, we obtain the curve for pure $\sprm(N)_0$ theory with theta angle zero.
\paragraph{\underline{$\sprm(N)_{\pi}$}}
The curve for $\sprm(N)_{\pi}$ can be obtained from the curve of $\sprm(N)+\mathsf{F}$ by changing the sign of the mass parameter 
\begin{align}
    m_1\rightarrow -m_1,
\end{align}
and then take the massive limit $m_1\rightarrow \infty$. We can derive the curve as
\begin{align}\label{eq:curvepi}
    Y+\frac{1}{\sh(2x+\epsilon_1\pm \epsilon_1)\sh(2x+\epsilon_1)^2}Y^{-1}+V_0^{\prime\prime\prime}(x) =(-1)^{N-1}e^{-\frac{1}{2}m_0} \sum_{j=1}^{N}(-1)^{j+1}\ch(2x)^{N-j} H_j,
\end{align}
where 
\begin{align}
    V_0^{\prime\prime\prime}(x)=-(-1)^N\frac{1}{2\,\ch(x\pm \epsilon_1/2)}+\frac{\delta}{2\,\sh(x\pm \epsilon_1/2)}-e^{-\frac{1}{2}m_0}\ch(2x)^{N},
\end{align}
with $\delta=-1$. Specifically, when $\delta=1$, the curve is the curve for $\sprm(N)_{0}$.
\subsection{Wilson loops and quantum periods}\label{sec:TSTandWilson}
For a toric Calabi-Yau threefold, the corresponding quantum curve of a 5d $\mathcal{N}=1$ theory is expected to be the quantum mirror curve in the B-model \cite{Aganagic:2011mi, Huang:2014nwa}, and the Wilson loop expectation values are expected to be the complex structure parameters in B-model \cite{Huang:2022hdo}. Following the strategy in \cite{Aganagic:2011mi}, we can solve the quantum periods from the curve, afterward compute the inverse series and compare with the Wilson loop calculations described in \cite{Gaiotto:2015una} for $\sprm(N)$ theories as a check of the quantum curve. Similar check work for 5d $SU(N)$ cases has been done in \cite{Grassi:2018bci}. For the $\sprm(N)$ gauge group, the corresponding Calabi-Yau threefold is generally non-toric, and there is no direct B-model quantum mirror curve description at this moment. Even though, we can still conjecture that our quantum curve derived in the previous section describes the quantum mirror curve and the eigenvalues of the Hamiltonians $H_i$ are mapped to the B-model complex structure parameters $z_i$ via
\begin{align}\label{eq:Hizi}
    H_j =\prod_{i=1}^{N}z_i^{-C^{-1}_{ij}},\quad j=1,\cdots N,
\end{align}
and all other mass parameters $m_{i>0}$ as additional independent complex structure parameters. Here $C_{ij}$ is the Cartan matrix of the $\sprm(N)$ group and $-C_{ij}$ is the intersection matrix between compact divisors and curves.
In the definition \eqref{eq:Hizi}, the $H_j$ parameters are complex structure parameters that dual to the compact divisors, so they generate the Wilson loop of orbits instead of representations from ``lowest" weights which comes from the lowest degrees of the large K\"ahler parameter expansions of $H_j$. From the description here, we can determine the expression on the right-hand side of \eqref{eq:curve5} from the perturbative contribution of the codimension four defect partition function, by reading the lowest degrees in the large K\"ahler parameter expansions.

In the remaining part of this section, we will calculate the quantum periods for some rank-one and rank-two models and compare them with the results from the Wilson loop calculations.

\subsection{$\sprm(1)+7\mathsf{F}$}
The quantum curve for $\sprm(1)+7\mathsf{F}$ theory can be read from \eqref{eq:curve5} and under the redefinition of Hamiltonians, we have
\begin{align}
   H_1= Y+\frac{\prod_{l=1}^{8}\sh(x\pm m_l+\epsilon_1/2)}{\sh(2x+\epsilon_1\pm \epsilon_1)\sh(2x+\epsilon_1)^2}Y^{-1}+V_0(x),
\end{align}
where $V_0(x)$ can be read from \eqref{eq:V05}, but in order to have an enhanced $E_8$ global symmetry, we shift one of the mass parameters by a phase
\begin{align}
    m_8\rightarrow m_8+ i\pi.
\end{align}
Such that
\begin{align}
    V_0(x)=&-\frac{\prod_{l=1}^{8}\ch(m_l)}{2\,\ch(x\pm \epsilon_1/2)}-\frac{\prod_{l=1}^{8}\sh(m_l)}{2\,\sh(x\pm \epsilon_1/2)}-\ch(2x)\sum_{l=1}^{8}\ch(2m_l)+\ch(\epsilon_1)\,\ch(4x).
\end{align}
The curve has manifest $SO(16)$ symmetry
\begin{align}
    m_i\rightarrow -m_i,\quad i=1,\cdots, 8,
\end{align}
so that we can use $SO(16)$ characters to rewrite the curve. Denota $\chi_i$ the character of $\sorm(16)$ whose highest weight is the $i$-th highest weight labeled in \eqref{dynkin:D8},
\begin{equation}\label{dynkin:D8}
    \dynkin[%
edge length=.95cm,
labels*={1,...,8}]D{8} 
\end{equation}
we have the replacements
\begin{align}
    &\prod_{l=1}^{8}\sh(x\pm m_l+\epsilon_1/2)=\sum_{i=0}^6(-1)^{8-i}\ch((8-i)(2x+\epsilon_1))\chi_i\nonumber\\
    &\quad\quad+\ch(2x+\epsilon_1)(\chi_1+\chi_3+\chi_5-\chi_7\chi_8)+\chi_7^2+\chi_8^2-2(\chi_0+\chi_2+\chi_4+\chi_6)
\end{align}
and
\begin{align}
    \prod_{l=1}^{8}\ch(m_l)=\chi_7+\chi_8,\quad\quad \prod_{l=1}^{8}\sh(m_l)=\chi_7-\chi_8,\quad\quad \sum_{l=1}^8 \ch(m_l)=\chi_1,
\end{align}
such that we verify that our quantum curve in the classical limit $q\rightarrow 1$ agrees with the classical Seiberg-Witten curve from brane diagrams in \cite{Hayashi:2017btw, Li:2021rqr}. See \cite{Moriyama:2020lyk} for another quantum curve approach from generalized toric diagrams.
One can further compute the quantum periods of the curve by using the method in \cite{Aganagic:2011mi} and we find that the eigenvalue of the Hamiltonian has an enhanced $E_8$ global symmetry
\begin{align}
    H_1=\frac{1}{Q}+\left(7+q^2+q^{-2}+3\chi_{\mathbf{248}}+(q+q^{-1})(3+\chi_{\mathbf{248}})+\chi_{\mathbf{3875}}\right)Q+\mathcal{O}(Q^2),
\end{align}
where $\chi_{{\textbf{dim}}}$ is the character of $E_8$ with dimension $\textbf{dim}$.
\subsection{$\sprm(2)_0$}
The quantum curve of $\sprm(2)_0$ theory is 
\begin{align}\label{curve:SP2_0}
    Y+\frac{q^2 X^4}{(1-X^2)(1-q X^2)^2(1-q^2X^2)}Y^{-1} &+\frac{q^{-\frac{1}{2}}(1+q)X^2}{(1-q^{-1}X^2)(1-q X^2)}\nonumber\\
    &+q_0^{-1}(X+X^{-1})^2= q_0^{-1}H_1(X^{-1}+X)-q_0^{-1}H_2,
\end{align}
where $q_0=-e^{\frac{1}{2}m_0}$ is the instanton counting parameter. There are two independent A-periods, classically their independent components can be computed from the residues
\begin{align}
    \Pi_1=-\mathrm{Res}_{X\rightarrow 0} \frac{\log Y}{X},\quad \Pi_2=-\mathrm{Res}_{Y\rightarrow 0} \frac{\log X}{Y}. 
\end{align}
In topological string theory, the quantum curve is the quantized version of the mirror curve in the B-model, where the moduli space is described by the complex structure parameters. According to the description in \cite{Huang:2022hdo}, the complex structure parameters $z_1,z_2,z_3$ naturally connect to the Wilson loops of orbits rather than representations via the inverse of Cartan matrix as
\begin{align}
    H_1=\frac{1}{z_1\sqrt{z_2}},\quad \quad H_2=\frac{1}{z_1z_2}, \quad\quad q_0=z_3,
\end{align}
where
\begin{align}
    H_1=\lim_{\epsilon_2\rightarrow 0}\widetilde{W}_{\text{fund}},\quad\quad H_2=\lim_{\epsilon_2\rightarrow 0}\widetilde{W}_{\mathsf{\Lambda}^2}-1.
\end{align}
Around the small complex structures region $z_i\sim 0$, we can solve that
\begin{align}
    \Pi_1&=\log(z_1z_2z_3)+z_2+\frac{1}{2}(-4z_1z_2+3z_2^2)-\frac{2}{3}(9z_1z_2^2-5z_2^3+3z_1z_2^2z_3)+\cdots,\\
    \Pi_2&=\frac{1}{2}\log z_2+(-z_1+z_2)+\frac{1}{2}(-3z_1^2-4z_1z_2+3z_2^2)\nonumber\\
   & \quad\quad\quad\quad +\frac{1}{3}(-10z_1^3-3z_1^2z_2-18z_1z_2^2+10z_2^3-6z_1z_2^2z_3)+\cdots.
\end{align}
Identify the periods as Coulomb parameters
\begin{align}
    e^{\Pi_1}=q_0e^{\alpha_1+\alpha_2},\quad e^{\Pi_2}=e^{\frac{1}{2}\alpha_2},
\end{align}
we can test that the results agree with the Wilson loop expectation values calculated from 5d gauge theory in \cite{Gaiotto:2015una}. 
Quantum period calculation gives
\begin{align}
    \Pi_1(q)&=\log(z_1z_2z_3)+z_2+\frac{1}{2}(-4z_1z_2+3z_2^2)+\left(-6z_1z_2^2+\frac{10}{3}z_2^3-(q^{1/2}+q^{-1/2})z_1z_2^2z_3\right)\nonumber\\&+\left(\frac{35 z_2^4}{4}-20 z_1 z_2^3+3 z_1^2 z_2^2+(q^{1/2}+q^{-1/2})(z_1^2 z_2^2 z_3-5 z_1 z_2^3 z_3)-(q^{3/2}+q^{-3/2})z_1 z_2^3 z_3\right)+\cdots \,.
\end{align}
\subsection{$\sprm(2)_{\pi}$}
The quantum curve of $\sprm(2)_{\pi}$ theory is 
\begin{align}
    Y+\frac{q^2 X^4}{(1-X^2)(1-q X^2)^2(1-q^2X^2)}Y^{-1} &+\frac{X(1+X^2)}{(1-q^{-1}X^2)(1-q X^2)}\nonumber\\
    &-q_0^{-1}(X+X^{-1})^2= -q_0^{-1}H_1(X^{-1}+X)+q_0^{-1}H_2,
\end{align}
where $q_0=e^{\frac{1}{2}m_0}$ is the instanton counting parameter.
\begin{align}
    \Pi_1(q)&=\log(z_1z_2z_3)+z_2+\frac{1}{2}(-4z_1z_2+3z_2^2)+z_1z_2^{3/2}z_3+\left(-6z_1z_2^2+\frac{10}{3}z_2^3\right)\nonumber\\&+(q+4+q^{-1})4z_1z_2^{5/2}z_3+\left(\frac{35 z_2^4}{4}-20 z_1 z_2^3+3 z_1^2 z_2^2\right)+\cdots\,.
\end{align}
\subsection{$\mathbb{P}^2\cup\mathbb{F}_6$}
\label{4.6}
The theory $\mathbb{P}^2\cup\mathbb{F}_6$ is named in \cite{Jefferson:2018irk} as a non-Lagrangian theory that flows from the gauge theory $\sprm(2)_0$. It has a geometric description that comes from the gluing of two del Pezzo surfaces $\mathbb{P}^2$ and $\mathbb{F}_6$. By looking at the intersection numbers of the geometric description, one can determine that by taking the limit 
\begin{align}
    z_2\rightarrow \Lambda \,z_2,\quad\quad q_0\rightarrow \Lambda^{-2},\quad\quad \Lambda \rightarrow 0,
\end{align}
in the B-model description of the $\sprm(2)_0$ theory, we can obtain the theory of $\mathbb{P}^2\cup\mathbb{F}_6$. At the level of curve, we also need to do the shift in coordinates $X,Y$ with the scaling parameter $\Lambda$ to keep the dynamic information. In total, we have
\begin{align}\label{map:nonL}
    z_2\rightarrow \Lambda \,z_2,\quad\quad q_0\rightarrow \Lambda^{-2},\quad\quad Y\rightarrow \Lambda \, Y\quad\quad X\rightarrow \Lambda^{\frac{1}{2}}\,X,\quad\quad \Lambda \rightarrow 0.
\end{align}
By applying \eqref{map:nonL} to \eqref{curve:SP2_0}, we obtain the quantum curve of $\mathbb{P}^2\cup\mathbb{F}_6$ as 
\begin{align}
    Y+\frac{q^2 X^4}{Y} &+{q^{-\frac{1}{2}}(1+q)X^2}+X^{-2}= H_1X^{-1}-H_2.
\end{align}
The intersection matrix becomes
\begin{align}
    -C=\left(
\begin{array}{cc}
 -2 & 1 \\
 2 & -3 \\
\end{array}
\right)
\end{align}
which leads to the identification of the complex structure parameters
\begin{align}
    H_1=z_1^{-\frac{3}{4}}z_2^{-\frac{1}{4}},\quad\quad H_2=z_1^{-\frac{1}{2}}z_2^{-\frac{1}{2}}.
\end{align}
By using the method described in the previous sections, we compute the quantum periods
\begin{align}
    \Pi_1(q)=&\log(z_1)+\left(\sqrt{q} z_2+\frac{z_2}{\sqrt{q}}+2 z_1\right)+\left(-q^2 z_2^2-\frac{z_2^2}{q^2}-\sqrt{q} z_2 z_1-\frac{z_2 z_1}{\sqrt{q}}-\frac{7 q
   z_2^2}{2}-\frac{7 z_2^2}{2 q}\right.\nonumber\\
   &+3 z_1^2-6 z_2^2\bigg)+\bigg(q^{9/2} z_2^3+3 q^{7/2} z_2^3+12 q^{5/2} z_2^3+\frac{88}{3} q^{3/2} z_2^3+\frac{88
   z_2^3}{3 q^{3/2}}+\frac{12 z_2^3}{q^{5/2}}+\frac{3
   z_2^3}{q^{7/2}}\nonumber\\
   &+\frac{z_2^3}{q^{9/2}}+3 q^2 z_1 z_2^2+\frac{3 z_1 z_2^2}{q^2}+8 q z_1
   z_2^2+48 \sqrt{q} z_2^3+\frac{48 z_2^3}{\sqrt{q}}+\frac{8 z_1 z_2^2}{q}+\frac{20
   z_1^3}{3}+14 z_1 z_2^2\bigg)+\cdots,\nonumber\\
   \Pi_2(q)=&\log \left(z_2\right)+ \left(-3 \sqrt{q} z_2-\frac{3 z_2}{\sqrt{q}}-2 z_1\right)+
   \bigg(3 q^2 z_2^2+\frac{3 z_2^2}{q^2}+3 \sqrt{q} z_2 z_1+\frac{3 z_2
   z_1}{\sqrt{q}}+\frac{21 q z_2^2}{2}+\frac{21 z_2^2}{2 q}\nonumber\\
   &-3 z_1^2+18 z_2^2\bigg)+
   \bigg(-3 q^{9/2} z_2^3-9 q^{7/2} z_2^3-36 q^{5/2} z_2^3-88 q^{3/2} z_2^3-\frac{88
   z_2^3}{q^{3/2}}-\frac{36 z_2^3}{q^{5/2}}-\frac{9 z_2^3}{q^{7/2}}\nonumber\\
   &-\frac{3
   z_2^3}{q^{9/2}}-9 q^2 z_1 z_2^2-\frac{9 z_1 z_2^2}{q^2}-24 q z_1 z_2^2-144 \sqrt{q}
   z_2^3-\frac{144 z_2^3}{\sqrt{q}}-\frac{24 z_1 z_2^2}{q}-\frac{20 z_1^3}{3}-42 z_1
   z_2^2\bigg)+\cdots,\nonumber
\end{align}
which give the quantum mirror maps to the Coulomb parameters.

We in addition remark that one can similarly obtain the quantum curves of other 5d $\mathcal N=1$ non-Lagrangian theories so long as they sit in the higgsing trees of the $\sprm(N)$ theories. We omit the computation for brevity.

\subsection{$\sprm(2)+9\mathsf{F}$}
In the last example, we test the quantum curve of $\sprm(2)+9\mathsf{F}$ theory by computing one of the quantum A-periods similar to previous cases and then compare it with the result from Wilson loops and we find an exact agreement. Moreover, if we absorb the factor $e^{-\frac{1}{2}m_0}$ to the Hamiltonians $H_1=\frac{1}{z_1\sqrt{z_2}}$ and $H_2=\frac{1}{z_1z_2}$, then quantum period has manifest $SO(20)$ global symmetry
\begin{align}
    \Pi_1(q)&=\log(z_1z_2)+z_2+\frac{1}{2}(-4\chi_1 z_1z_2+3z_2^2)+z_1z_2^{3/2}(-\chi_c+2q^{3/2}+2q^{-3/2})\nonumber\\
    &+\left(((q^{1/2}+q^{-1/2})\chi_s-6\chi_1)z_1z_2^2+\frac{10}{3}z_2^3\right)+\cdots,
\end{align}
where $\chi_1$, $\chi_s$ and $\chi_c$ are the $\sorm(20)$ characters for fundamental, spinor and conjugate spinor representations.

\section{Conclusion}
In this paper, we investigate the 6d D-type minimal conformal matter theory in the presence of codimension two and four surface defects, also known as Wilson surface defects. These two types of defects play important roles in the quantization of Seiberg-Witten curves of the 6d theory compactified on $\mathbb R^4\times\mathbb T^2$. Specifically, we demonstrate that the BPS instanton partition function with the insertion of the codimension two defect acts as an eigenfunction of the quantized Seiberg-Witten curve, while the Wilson surface defect acts as the eigenvalue of the quantum curve. Our results extend previous findings on E-string theory, where the quantum curve was identified as the van Diejen difference operator in the integrability community \cite{Chen:2021ivd}. Along this line, we show that the quantum curve of D-type minimal conformal matter can be identified with a type of Elliptic Garnier system. Moreover, when the 6d theory is compactified onto a circle, we obtain the corresponding 5d KK theory, $\sprm(N)$ with $2N+6$ flavors. By taking the masses of flavors to infinity, we showed that the 6d quantum curve can be deformed into a series of quantum curves associated with 5d $\sprm(N)$ theory with matters $N_f\leq 2N+5$. 

Our current work opens up several avenues for further research. Firstly, one can study the SW-curves for A-type quiver theories that consist of single-node 6d SCFTs defined on $-n$ curves. One such example is the linear tensor chain, which is realized on $-1$ and $-4$ curves and corresponds to the D-type conformal matters. The quantum curve of the $\sorm(N)$ theory on the $-4$ curve has already been obtained in \cite{Chen:2021rek}. Another interesting linear quiver is the higher rank E-string theory and its generalizations \cite{Mekareeya:2017jgc}. One particularly interesting feature of these quiver theories is that they can be Higgsed to various 6d SCFTs. Therefore, starting with SW-curves of the quiver theories, we can investigate many interesting non-perturbative data of the Higged theories as well as their own quantum curves. On the other hand, one can also compactify the quiver theories onto a circle and study their deformations to 5d SCFTs and associated quantum curves as we have done in the note. We would thus be able to obtain many ``quantum curve cascades" from both Higgsings and deformations.

Another interesting direction is to systematically study the relations between the 6d SW-curves and elliptic integrable systems. A series of investigations have demonstrated that the quantum curves in several 6d SCFTs can be identified to various elliptic integrable systems \cite{Bullimore:2014awa, Chen:2020jla, Chen:2021ivd, Chen:2021rek}. It is then natural to expect the correspondence held for generic cases. For example, in the note, we have established the relation between quantum curves of D-type minimal conformal matters and elliptic Garnier systems. A further generalization could be the quantum curves of higher rank E-string and the $BC_{N}$ system. In addition, for a quiver-like 6d SCFT, a notable distinction from the single-node ones is the much richer varieties of codimension two and four defects that can be introduced. It would be intriguing to understand how these surface defects could engineer the quantum curves and their roles in the associated elliptic integrable systems. Moreover, we have also shown in this note that the mass deformations of 6d quantum curves could result in a cascade of quantum curves. Therefore an immediate interesting question raised is to interpret these deformations also from the perspective of integrable systems. Such interpretation has been investigated between various 4d Seiberg-Witten curves of $\surm(2)+N_f\mathsf{F}$ and Argyres-Douglas theories, and the isomonodromic deformations of linear differential systems \cite{Bonelli:2016qwg}. It would be fascinating to establish a similar correspondence between the deformations of 6d SW-curves and their analogs in the field of elliptic integrable systems. That is another exciting topic we hope to explore further in the future.

\paragraph{Acknowledgments}

We thank Babak Haghighat, Hee-Cheol Kim, Kimyeong Lee and Marcus Sperling for their collaborations on previous subjects. We also thank Oleg Chalykh, Sung-Soo Kim, Yuji Sujimoto, Futoshi Yagi for their helpful discussions. The work of JC is supported by the Fundamental Research Funds for the Central Universities (No.20720230010) of China. YL and XW are supported by KIAS Individual Grants PG084801 and QP079202.
\clearpage
\appendix
\section{Useful identities}\label{A}
 In this appendix, we collect some useful theta function identities used in the paper.
 We use $\vartheta_i(z;\tau)$ for $i=1,2,3,4$ to denote the Jaboci theta functions. In the elliptic genus , we also use a useful variant of Jaboci theta functions denoted as 
 \begin{align}
     \theta_i (z;\tau) = \frac{\vartheta_i(z;\tau)}{i\eta(\tau)},
 \end{align}
 where $\eta(\tau)$ is the Dedekind eta function. For convenience, we often omit to indicate explicitly the dependence on $\tau$. 

 We also use the $p$-theta function
 \begin{align}
    \left[X\right]= 
    \prod_{k=0}^{\infty} (1-p^{k} X) (1-p^{k+1} /X)= \PE\bigg[\frac{X + p/X}{(1-p)}\bigg]
    = X^{\frac{1}{2}} p^{-\frac{1}{12}} \theta_1(\frac{x}{2 \pi i}),
 \end{align}
 where we have used the multiplicative variables $X = e^{2\pi i x}$ and $p =e^{2\pi i \tau}$, and ${\rm PE}$ is plethystic exponential function defined as 
 \begin{align}
{\rm PE}[f(z)] = \exp \left (\displaystyle\sum_{k=1}^{\infty} \frac{f(z^k)}{k} \right).
 \end{align} 
 $\left[X\right]$ satisfies the functional relation
 \begin{align}
     \left[X\right] \equiv \frac{\Gamma_{p,q}(q X)}{\Gamma_{p,q}(X)},
 \end{align}
 where $\Gamma_{p,q}(X)$ the elliptic Gamma functions defined by
 \begin{align}
     \Gamma_{p,q}(X) = \prod_{i,j =0 }^{\infty} \frac{1-p^{i+1}q^{j+1} /X}{1-p^{i}q^{j} X} = \PE\bigg[\frac{X- p q/X}{(1-p)(1- q)}\bigg].
 \end{align}
We use the shorthanded notation 
\begin{align}
    \vartheta_b^{\Delta}(x) := \frac{\vartheta_b'(x - \frac{\epsilon_1}{2})}{\vartheta_b(x - \frac{\epsilon_1}{2})} -  \frac{\vartheta_b'(x + \frac{\epsilon_1}{2})}{\vartheta_b(x + \frac{\epsilon_1}{2})}. \label{eq:thetadelta}
\end{align}
We derived two sets of identities for the theta functions of degree 2: 
\begin{align}
   & \vartheta_b^{\Delta}(x) = \frac{1}{\vartheta_c(\frac{\epsilon_1}{2})^2}\left((-1)^{\alpha_{a,b}} \frac{\vartheta_1'(0) \vartheta_1(\epsilon_1) \vartheta_a(x)^2}{\vartheta_b(x \pm \frac{\epsilon_1}{2})} - 2 \vartheta_c'(\frac{\epsilon_1}{2})\vartheta_c(\frac{\epsilon_1}{2}) \right), \\
   & \vartheta_1(x \pm \alpha)  = \frac{1}{\vartheta_c(\frac{\epsilon_1}{2})^2} \left( (-1)^{\beta_{a,b}} \vartheta_a(x)^2 \vartheta_b (\frac{\epsilon_1}{2} \pm \alpha) + (-1)^{\gamma_{a,b}} \vartheta_a(\alpha)^2 \vartheta_b(x \pm \frac{\epsilon_1}{2})\right),
\end{align}
where $\omega_c = \omega_a + \omega_b$ (mod $\mathbb{Z} + \tau \mathbb{Z}$) with $a, b, c $ runs from 1 to 4, 
and 
    \begin{align}
        \alpha_{a,b}=  \begin{cases}
             1, \quad \text{for}\, (a, b) \in A, \\
             0, \quad \text{otherwise},
         \end{cases}
         \beta_{a,b}= \begin{cases}
             1, \quad \text{for}\, (a, b) \in B, \\
             0, \quad \text{otherwise},
         \end{cases}
      \gamma_{a,b}=  \begin{cases}
             0, \quad \text{for}\, (a, b) \in 
             B \cup C, \\
             1, \quad \text{otherwise},
         \end{cases}
    \end{align}
with $A = \{(1,3), (3,4), (4,3), (2,4)\}$, $B=\{(3,3), (2,3),(3,4),(2,4)\}$ and $C =\{(a,1) \mid a =2,3,4\}$. For The cases with $a \neq b$ can be viewed as the quantum version of addition formulae.

We also need to use the following addition formulae
\begin{align} \label{addform}
    \vartheta_a(x \pm \frac{\epsilon_1}{2}) = \frac{1}{\vartheta_4(0)^2} \left(\vartheta_b(\frac{\epsilon_1}{2})^2 \vartheta_1(x)^2 -\vartheta_a(\frac{\epsilon_1}{2})^2  \vartheta_4(x)^2  \right),
\end{align}
where $b$ satisfied $\omega_b = \omega_a + \omega_4 $ (mod $\mathbb{Z} + \tau \mathbb{Z}$). 

The following identities can be used to express the square of any theta function in terms of the squares of any two other theta functions 
\begin{align} \label{squares}
    \vartheta_3(0)^2 \vartheta_3(x)^2 & =  \vartheta_4(0)^2 \vartheta_4(x)^2 +  \vartheta_2(0)^2 \vartheta_2(x)^2\,, \notag \\
    \vartheta_3(0)^2 \vartheta_4(x)^2 & =  \vartheta_2(0)^2 \vartheta_1(x)^2 +  \vartheta_4(0)^2 \vartheta_3(x)^2\,, \notag \\
    \vartheta_2(0)^2 \vartheta_4(x)^2 & =  \vartheta_3(0)^2 \vartheta_1(x)^2 +  \vartheta_4(0)^2 \vartheta_2(x)^2\,, \notag \\
    \vartheta_2(0)^2 \vartheta_3(x)^2 & =  \vartheta_4(0)^2 \vartheta_1(x)^2 +  \vartheta_3(0)^2 \vartheta_2(x)^2 \,.
\end{align}

\section{Two-instanton results under NS-limit}
\label{B}
In this appendix, we collect the two-instanton results in the presence of codimension two and four defects. It will involve $\Psi_2(x;\epsilon_1)$ and $\chi_2(x;\epsilon_1)$ which are defined as
\begin{align}
 \Psi_2(x;\epsilon_1)= & \lim_{\epsilon_2\rightarrow 0}\left(Z^{\rm 6d/4d}_2-Z^{\rm 6d}_2-Z^{\rm 6d}_1\left(Z^{\rm 6d/4d}_1-Z^{\rm 6d}_1\right)\right) \,,\notag\\*
\chi_2(x;\epsilon_1) = & \lim_{\epsilon_2\rightarrow 0}\left( Z^{\rm 6d/2d}_2-Z^{\rm 6d}_2-Z^{\rm 6d }_1\left(Z^{\rm 6d/2d}_1-Z^{\rm 6d}_1\right) \right) \,.
\end{align}
where the two-instanton partition function with a 4d defect $Z_2^{\text{4d-6d}}$ is given by
\begin{align}
Z_2^{{\rm 6d/4d}}=\frac{1}{2}Z^{{\rm 6d/4d}}_{2,\, {\rm cont.}}+\frac{1}{4}\sum_{a=2}^4 Z^{\rm 6d/ 4d}_{2(a),\,{\rm dis.}}\,,
\label{eq:2-inst_defect}
\end{align}
with
\begin{align}
Z^{\rm 6d/4d}_{2(a),\,{\rm dis.}}\,= \, & 
 \frac{{\vartheta}_a(0) {\vartheta}_a(2\epsilon_+)}{\eta^{12}{\vartheta}_1(\epsilon_{1,2})^2
 {\vartheta}_a(\epsilon_{1,2})}
 \,  \bigg(\frac{  \prod_{f=1}^{2N+8}  \vartheta_1(m_f ) \vartheta_a(m_f) }{\prod_{i=1}^N\vartheta_1(\pm\alpha_i+\epsilon_+)\vartheta_a(\pm\alpha_i+\epsilon_+)}
\, \frac{\vartheta_1(x+\epsilon_+)\vartheta_a(x+\epsilon_+)}{\vartheta_1(x+\epsilon_-)\vartheta_a(x+\epsilon_-)}
\notag\\
&\quad 
+ \prod_{i=1}^2 \frac{  \prod_{f=1}^{2N+8}  \vartheta_{\sigma^i(a)}(m_f )   }{\prod_{i=1}^N\vartheta_{\sigma^i(a)}(\pm\alpha_i+\epsilon_+)}
\, \frac{\vartheta_{\sigma^i(a)}(x+\epsilon_+)}{\vartheta^i_{\sigma(a)}(x+\epsilon_-)}\bigg)\,,
 \end{align}
and
\begin{align}
Z_{2,\,{\rm cont.}}^{\rm 6d/4d} =&
  \frac{-1}{{2}\eta^{12}
  \vartheta_1(\epsilon_{1,2}) \vartheta_1(2\epsilon_1)\vartheta_1(2\epsilon_-)}
  \sum_{a=1}^4
\bigg(
 \frac{\prod_{f=1}^{2N+8}
 \vartheta_a(m_f \pm {\epsilon_{1}}/{2} )
 }{\prod_{i=1}^N\vartheta_a(\pm\alpha_i+\epsilon_+\pm\epsilon_1/2)}
\frac{\vartheta_a(x+\epsilon_2/2)\vartheta_a(x+\epsilon_1+\epsilon_2/2)}{\vartheta_a(x-\epsilon_2/2)\vartheta_a(x+\epsilon_1-\epsilon_2/2)} 
\bigg)\notag\\
&+\frac{1}{{2}\eta^{12}
  \vartheta_1(\epsilon_{1,2}) \vartheta_1(2\epsilon_2)\vartheta_1(2\epsilon_-)}
 \sum_{a=1}^4
\bigg(
 \frac{\prod_{f=1}^{2N+8}
 \vartheta_a(m_f \pm {\epsilon_{2}}/{2} )
 }{\prod_{i=1}^N\vartheta_a(\pm\alpha_i+\epsilon_+\pm\epsilon_2/2)}
 \frac{\vartheta_a(x+\epsilon_2+\epsilon_1/2)}{\vartheta_a(x-\epsilon_2+\epsilon_1/2)} 
\bigg)\notag\\
 &+\sum_{i=1}^N\bigg(\frac{1}{\eta^{12}\vartheta_1(\epsilon_{1,2})\vartheta_1(2\alpha_i)\vartheta_1(2\epsilon_+-2\alpha_i)\vartheta_1(2\alpha_i-\epsilon_{1,2})\vartheta_1(-2\alpha_i+2\epsilon_++\epsilon_{1,2})}\notag\\ 
 &
 \quad \frac{\prod_{f=1}^{2N+8}\vartheta_1(\alpha_i\pm m_f-\epsilon_+)}{\prod_{j\neq i}^N\vartheta_1(\alpha_i\pm\alpha_j)\vartheta_1(\alpha_i\pm\alpha_j-2\epsilon_+)} \frac{\vartheta_1(\alpha_i+x)\vartheta_1(-\alpha_i+x+2\epsilon_+)}{\vartheta_1(\alpha_i+x-\epsilon_2)\vartheta_1(-\alpha_i+x+\epsilon_1)} 
+(\alpha_i\rightarrow-\alpha_i) \bigg)\notag\\
&+\frac{\vartheta_1(2\epsilon_+)}{\eta^{12}\vartheta_1(\epsilon_1)\vartheta_1(2x+2\epsilon_-)\vartheta_1(2x+2\epsilon_1-\epsilon_2)\vartheta_1(2x-\epsilon_2)\vartheta_1(2x+\epsilon_1-2\epsilon_2)}\notag\\
&\quad \frac{\prod_{f=1}^{2N+8}\vartheta_1(x\pm m_f+\epsilon_-)}{\prod_{i=1}^N\vartheta_1(2x\pm\alpha_i+\epsilon_1)\vartheta_1(2x\pm\alpha_i-\epsilon_2)}\,,
\end{align}
and the two-instanton partition function with a 2d defect $Z_2^{\rm 6d/2d}$ is given by 
\begin{align}
Z_2^{\rm 6d/2d}=\frac{1}{2} Z^{\rm 6d/2d}_{2,\, {\rm cont.}}+\frac{1}{4}\sum_{a=2}^4 Z^{\rm 6d/2d}_{2(a),\,{\rm dis.}}\,,
\label{eq:2-inst_Wilson}
\end{align}
with
\begin{align}
Z^{\rm 6d/2d}_{2(a),\,{\rm dis.}}&= 
 \frac{\vartheta_a(0) \vartheta_a(2\epsilon_+)}{\eta^{12}\vartheta_1(\epsilon_{1,2})^2
 \vartheta_a(\epsilon_{1,2})}
  \bigg(\frac{  \prod_{f=1}^{2N+8}  \vartheta_1(m_f ) \vartheta_a(m_f) }{\prod_{i=1}^N\vartheta_1(\pm\alpha_i+\epsilon_+)\vartheta_a(\pm\alpha_i+\epsilon_+)}
 \frac{\vartheta_1(x\pm\epsilon_-)\vartheta_a(x\pm\epsilon_-)}{\vartheta_1(x\pm\epsilon_+)\vartheta_a(x\pm\epsilon_+)}\notag\\
&+\frac{  \prod_{f=1}^{2N+8}  \vartheta_{\sigma(a)}(m_f ) \vartheta_{\sigma^2(a)}(m_f) }{\prod_{i=1}^N\vartheta_{\sigma(a)}(\pm\alpha_i+\epsilon_+)\vartheta_{\sigma^2(a)}(\pm\alpha_i+\epsilon_+)}
\frac{\vartheta_{\sigma(a)}(x\pm\epsilon_-)\vartheta_{\sigma^2(a)}(x\pm\epsilon_-)}{\vartheta_{\sigma(a)}(x\pm\epsilon_+)\vartheta_{\sigma^2(a)}(x\pm\epsilon_+)}
\bigg)\,,
\end{align}
and
\begin{align}
Z_{2,\,{\rm cont.}}^{\rm 6d/2d}&=
  \frac{-1}{{2}\eta^{12}
  \vartheta_1(\epsilon_{1,2}) \vartheta_1(2\epsilon_1)\vartheta_1(2\epsilon_-)}
   \sum_{a=1}^4
\Big(
 \frac{\prod_{f=1}^{2N+8}
 \vartheta_a(m_f \pm {\epsilon_{1}}/{2} )
 }{\prod_{i=1}^N\vartheta_a(\pm\alpha_i+\epsilon_+\pm\epsilon_1/2)}\frac{\vartheta_a(x\pm\left(\epsilon_1-\frac{\epsilon_2}{2}\right))}{\vartheta_a(x\pm\left(\epsilon_1+\frac{\epsilon_2}{2}\right))}
+(\epsilon_1\leftrightarrow\epsilon_2)\Big) \notag\\
 &+\sum_{i=1}^N\bigg(\frac{1}{\eta^{12}\vartheta_1(\epsilon_{1,2})\vartheta_1(2\alpha_i)\vartheta_1(2\epsilon_+-2\alpha_i)\vartheta_1(2\alpha_i-\epsilon_{1,2})\vartheta_1(-2\alpha_i+2\epsilon_+ +\epsilon_{1,2})}\notag\\ 
&\quad \frac{\prod_{f=1}^{2N+8}\vartheta_1(\alpha_i\pm m_f-\epsilon_+)}{\prod_{j\neq i}^N\vartheta_1(\alpha_i\pm\alpha_j)\vartheta_1(\alpha_i\pm\alpha_j-2\epsilon_+)}
 \frac{\vartheta_1(\alpha_i\pm x-\epsilon_{1,2})}{\vartheta_1(\alpha_i\pm x)\vartheta_1(\alpha_i\pm x-2\epsilon_+)}+(\alpha_i\rightarrow-\alpha_i) \bigg) \notag\\*
&+ \bigg( \frac{1}{\eta^{12}\vartheta_1(2x)\vartheta_1(2x+2\epsilon_+)\vartheta_1(2x+2\epsilon_++\epsilon_{1,2})}
\frac{\prod_{f=1}^{2N+8}
\vartheta_1(x\pm m_f+\epsilon_+)}{\prod_{i=1}^N\vartheta_1(x\pm\alpha_i)\vartheta_1(x\pm\alpha_i+2\epsilon_+)}
\,\notag\\
&\quad 
+(\epsilon_{1,2}\rightarrow -\epsilon_{1,2}) \bigg) \,.
\end{align}
Putting the pieces together, we list the results below:
\begin{description}
    \item 
 $$\Psi_2 := \Psi_2(x;\epsilon_1)\,,$$
{ \footnotesize
\begin{align}
\Psi_2 &=\frac{1}{4 \eta^{12} \vartheta_1^\prime(0) \vartheta_1(\pm 
\epsilon_1)\vartheta_1(2\epsilon_1)} 
\sum_{a=1}^4 
\left[
\frac{ \vartheta_a^\prime (x+\epsilon_1) }{
\vartheta_a(x+\epsilon_1) }
+
\frac{ \vartheta_a^\prime (x) }{
\vartheta_a(x) }
\right]
\frac{\prod_{f=1}^{2N+8} \vartheta_a(m_f\pm \epsilon_1/2)}{\prod_{i=1}^N\vartheta_a(\pm \alpha_i)\vartheta_a(\pm \alpha_1+\epsilon_1)}\notag\\
 &+ \frac{1}{8 \eta^{12} \vartheta_1^\prime(0)^2 \vartheta_1(\epsilon_1)^2} 
\sum_{a=1}^4 \frac{\vartheta_a^\prime\left(x+\frac{\epsilon_1}{2}\right)^2 }{ \vartheta_a\left(x+\frac{\epsilon_1}{2}\right)^2  }
 \cdot\frac{\prod_{f=1}^{2N+8} \vartheta_a(m_f)^2}{\prod_{i=1}^{N}\vartheta_a(\pm\alpha_i+\frac{\epsilon_1}{2})^2}\notag\\
&+ \frac{\vartheta_1^\prime(\epsilon_1)}{4 \eta^{12} 
\vartheta_1^\prime(0)^2 \vartheta_1(\epsilon_1)^3} \sum_{a=1}^4 \frac{\vartheta_a^\prime\left(x+\frac{\epsilon_1}{2}\right) }{ \vartheta_a\left(x+\frac{\epsilon_1}{2}\right)  }
 \cdot\frac{\prod_{f=1}^{2N+8} \vartheta_a(m_f)^2}{\prod_{i=1}^{N}\vartheta_a(\pm\alpha_i+\frac{\epsilon_1}{2})^2}\notag\\
&+ \frac{1}{4 \eta^{12} \vartheta_1^\prime(0)^2 \vartheta_1(\epsilon_1)^2} 
  \bigg[ 
  \frac{ \vartheta_1^\prime\left(x+\frac{\epsilon_1}{2}\right) \vartheta_2^\prime\left(x+\frac{\epsilon_1}{2}\right) }{ \vartheta_1\left(x+\frac{\epsilon_1}{2}\right)\vartheta_2\left(x+\frac{\epsilon_1}{2}\right) }
 \cdot\frac{\prod_{f=1}^{2N+8} \vartheta_1(m_f) \vartheta_2(m_f)}{\prod_{i=1}^N\vartheta_1(\pm\alpha_i+\frac{\epsilon_1}{2})\vartheta_2(\pm\alpha_i+\frac{\epsilon_1}{2})}\notag\\
 &\qquad\qquad\qquad\qquad+
  \frac{ \vartheta_1^\prime\left(x+\frac{\epsilon_1}{2}\right) \vartheta_3^\prime\left(x+\frac{\epsilon_1}{2}\right) }{ \vartheta_1\left(x+\frac{\epsilon_1}{2}\right) \vartheta_3\left(x+\frac{\epsilon_1}{2}\right)  }
 \cdot\frac{\prod_{f=1}^{2N+8} \vartheta_1(m_f) \vartheta_3(m_f)}{\prod_{i=1}^N\vartheta_1(\pm\alpha_i+\frac{\epsilon_1}{2})\vartheta_3(\pm\alpha_i+\frac{\epsilon_1}{2})}
 \notag \\
&\qquad \qquad \qquad\qquad     +
  \frac{ \vartheta_1^\prime\left(x+\frac{\epsilon_1}{2}\right) \vartheta_4^\prime\left(x+\frac{\epsilon_1}{2}\right) }{ \vartheta_1\left(x+\frac{\epsilon_1}{2}\right) \vartheta_4\left(x+\frac{\epsilon_1}{2}\right) }
\cdot\frac{\prod_{f=1}^{2N+8} \vartheta_1(m_f) \vartheta_4(m_f)}{\prod_{i=1}^N\vartheta_1(\pm\alpha_i+\frac{\epsilon_1}{2})\vartheta_4(\pm\alpha_i+\frac{\epsilon_1}{2})}\notag\\
&\qquad \qquad \qquad\qquad  
 +
  \frac{ \vartheta_2^\prime\left(x+\frac{\epsilon_1}{2}\right) \vartheta_3^\prime\left(x+\frac{\epsilon_1}{2}\right)}{ \vartheta_2\left(x+\frac{\epsilon_1}{2}\right)\vartheta_3\left(x+\frac{\epsilon_1}{2}\right)  }
 \cdot\frac{\prod_{f=1}^{2N+8} \vartheta_2(m_f) \vartheta_3(m_f)}{\prod_{i=1}^N\vartheta_2(\pm\alpha_i+\frac{\epsilon_1}{2})\vartheta_3(\pm\alpha_i+\frac{\epsilon_1}{2})}
 \notag \\
&\qquad \qquad \qquad\qquad     +
  \frac{ \vartheta_2^\prime\left(x+\frac{\epsilon_1}{2}\right) \vartheta_4^\prime\left(x+\frac{\epsilon_1}{2}\right) }{ \vartheta_2\left(x+\frac{\epsilon_1}{2}\right) \vartheta_4\left(x+\frac{\epsilon_1}{2}\right)  }
 \cdot\frac{\prod_{f=1}^{2N+8} \vartheta_2(m_f) \vartheta_4(m_f)}{\prod_{i=1}^N\vartheta_2(\pm\alpha_i+\frac{\epsilon_1}{2})\vartheta_4(\pm\alpha_i+\frac{\epsilon_1}{2})}\notag\\
&\qquad \qquad \qquad\qquad  
 +
  \frac{ \vartheta_3^\prime\left(x+\frac{\epsilon_1}{2}\right)\vartheta_4^\prime\left(x+\frac{\epsilon_1}{2}\right) }{ \vartheta_3\left(x+\frac{\epsilon_1}{2}\right)\vartheta_4\left(x+\frac{\epsilon_1}{2}\right)  }
\cdot\frac{\prod_{f=1}^{2N+8} \vartheta_3(m_f) \vartheta_4(m_f)}{\prod_{i=1}^N\vartheta_3(\pm\alpha_i+\frac{\epsilon_1}{2})\vartheta_4(\pm\alpha_i+\frac{\epsilon_1}{2})}\,
\bigg]\notag\\
&+
\frac{1}{ 4 \eta^{12} \vartheta_1^\prime(0)^2 \vartheta_1(\epsilon_1)^2}
\frac{\vartheta_2^\prime (\epsilon_1)}{\vartheta_2( \epsilon_1)}
\bigg[
  \left(\frac{ \vartheta_1^\prime \left(x+\frac{\epsilon_1}{2}\right)   }{ \vartheta_1\left(x+\frac{\epsilon_1}{2}\right) }
+
  \frac{ \vartheta_2^\prime \left(x+\frac{\epsilon_1}{2}\right)   }{ \vartheta_2\left(x+\frac{\epsilon_1}{2}\right) }\right)
\frac{\prod_{f=1}^{2N+8} \vartheta_1(m_f) \vartheta_2(m_f)}{\prod_{i=1}^{N}\vartheta_1(\pm\alpha_i+\frac{\epsilon_1}{2})\vartheta_2(\pm\alpha_i+\frac{\epsilon_1}{2})} 
\notag \\
&\qquad \qquad \qquad  \qquad\qquad \quad
 +
\left(\frac{ \vartheta_3^\prime \left(x+\frac{\epsilon_1}{2}\right)   }{ \vartheta_3\left(x+\frac{\epsilon_1}{2}\right) }
+
  \frac{ \vartheta_4^\prime \left(x+\frac{\epsilon_1}{2}\right)   }{ \vartheta_4\left(x+\frac{\epsilon_1}{2}\right) }\right)
\frac{\prod_{f=1}^{2N+8} \vartheta_3(m_f) \vartheta_4(m_f)}{\prod_{i=1}^{N}\vartheta_3(\pm\alpha_i+\frac{\epsilon_1}{2})\vartheta_4(\pm\alpha_i+\frac{\epsilon_1}{2})}
\bigg]
 \notag \\
&+
\frac{1}{ 4 \eta^{12} \vartheta_1^\prime(0)^2 \vartheta_1(\epsilon_1)^2}
\frac{\vartheta_3^\prime (\epsilon_1)}{\vartheta_3( \epsilon_1)}
\bigg[
  \left(\frac{ \vartheta_1^\prime \left(x+\frac{\epsilon_1}{2}\right)   }{ \vartheta_1\left(x+\frac{\epsilon_1}{2}\right) }
+
  \frac{ \vartheta_3^\prime \left(x+\frac{\epsilon_1}{2}\right)   }{ \vartheta_3\left(x+\frac{\epsilon_1}{2}\right) }\right)
\frac{\prod_{f=1}^{2N+8} \vartheta_1(m_f) \vartheta_3(m_f)}{\prod_{i=1}^{N}\vartheta_1(\pm\alpha_i+\frac{\epsilon_1}{2})\vartheta_3(\pm\alpha_i+\frac{\epsilon_1}{2})} 
\notag \\
&\qquad \qquad \qquad  \qquad\qquad \quad
 +
\left(\frac{ \vartheta_2^\prime \left(x+\frac{\epsilon_1}{2}\right)   }{ \vartheta_2\left(x+\frac{\epsilon_1}{2}\right) }
+
  \frac{ \vartheta_4^\prime \left(x+\frac{\epsilon_1}{2}\right)   }{ \vartheta_4\left(x+\frac{\epsilon_1}{2}\right) }\right)
\frac{\prod_{f=1}^{2N+8} \vartheta_2(m_f) \vartheta_4(m_f)}{\prod_{i=1}^{N}\vartheta_2(\pm\alpha_i+\frac{\epsilon_1}{2})\vartheta_4(\pm\alpha_i+\frac{\epsilon_1}{2})}
\bigg]
 \notag \\
&+
\frac{1}{ 4 \eta^{12} \vartheta_1^\prime(0)^2 \vartheta_1(\epsilon_1)^2}
\frac{\vartheta_4^\prime (\epsilon_1)}{\vartheta_4( \epsilon_1)}
\bigg[
  \left(\frac{ \vartheta_1^\prime \left(x+\frac{\epsilon_1}{2}\right)   }{ \vartheta_1\left(x+\frac{\epsilon_1}{2}\right) }
+
  \frac{ \vartheta_4^\prime \left(x+\frac{\epsilon_1}{2}\right)   }{ \vartheta_4\left(x+\frac{\epsilon_1}{2}\right) }\right)
\frac{\prod_{f=1}^{2N+8} \vartheta_1(m_f) \vartheta_4(m_f)}{\prod_{i=1}^{N}\vartheta_1(\pm\alpha_i+\frac{\epsilon_1}{2})\vartheta_4(\pm\alpha_i+\frac{\epsilon_1}{2})} 
\notag \\
&\qquad \qquad \qquad  \qquad\qquad \quad
 +
\left(\frac{ \vartheta_2^\prime \left(x+\frac{\epsilon_1}{2}\right)   }{ \vartheta_2\left(x+\frac{\epsilon_1}{2}\right) }
+
  \frac{ \vartheta_3^\prime \left(x+\frac{\epsilon_1}{2}\right)   }{ \vartheta_3\left(x+\frac{\epsilon_1}{2}\right) }\right)
\frac{\prod_{f=1}^{2N+8} \vartheta_2(m_f) \vartheta_3(m_f)}{\prod_{i=1}^{N}\vartheta_2(\pm\alpha_i+\frac{\epsilon_1}{2})\vartheta_3(\pm\alpha_i+\frac{\epsilon_1}{2})}
\bigg]
 \notag \\
&+\sum_{i=1}^N\bigg(\frac{1}{2\eta^{12}\vartheta_1(\epsilon_{1})\vartheta^\prime_1(0)\vartheta_1(2\alpha_i)^2\vartheta_1(2\alpha_i-\epsilon_1)^3\vartheta_1(2\alpha_i-2\epsilon_1)}\notag\\
&\quad\cdot\frac{\prod_{l=1}^{2N+8}\vartheta_1(\alpha_i\pm m_f-\frac{\epsilon_1}{2})}{\prod_{j\neq i}^N\vartheta_1(\alpha_i\pm\alpha_j)\vartheta_1(\alpha_i\pm\alpha_j-2\epsilon_1)}\cdot\left(\frac{\vartheta^\prime_1(x+\alpha_i)}{\vartheta_1(x+\alpha_i)}+\frac{\vartheta^\prime_1(x-\alpha_i+\epsilon_1)}{\vartheta_1(x-\alpha_i+\epsilon_1)}\right)\bigg)+(\alpha_i\rightarrow-\alpha_i)\notag\\
&+\frac{1}{2\eta^{12}\vartheta_1(2x+2\epsilon_1)\vartheta_1(2x+\epsilon_1)^2\vartheta_1(2x)}\cdot \frac{\prod_{f=1}^{2N+8}\vartheta_1(x\pm m_f+\frac{\epsilon_1}{2})}{\prod_{i=1}^N\vartheta_1(2x\pm\alpha_i+\epsilon_1)\vartheta_1(2x\pm\alpha_i)}\,.
\label{eq:2-inst_Psi}
\end{align}
}
\notag\\

\item $$\chi_2 := \chi_2(x;\epsilon_1)\,,$$ 
{\footnotesize
\begin{align}
\chi_2&=\lim_{\epsilon_2\rightarrow 0}\left(W^{\rm 6d/4d}_2-Z^{\rm 6d}_2-Z^{\rm 6d}_1\left(W^{\rm 6d/4d}_1-Z^{\rm 6d}_1\right)\right)\notag\\
&=\frac{1}{4\eta^{12}\vartheta_1(\epsilon_1)^2\vartheta_1(2\epsilon_1)\vartheta_1^\prime(0)}\sum_{a=1}^4\frac{\prod_{f=1}^{2N+8}\vartheta_a\left(m_f\pm\frac{\epsilon_1}{2}\right)}{\prod_{i}^N\vartheta_a(\pm\alpha_i)\vartheta_a(\pm\alpha_i+\epsilon_1)}\cdot\left(\frac{\vartheta_a^\prime(x+\epsilon_1)}{\vartheta_a(x+\epsilon_1)}-\frac{\vartheta_a^\prime(x-\epsilon_1)}{\vartheta_a(x-\epsilon_1)}\right)
\notag\\
&+\frac{1}{4\eta^{12}\vartheta_1(\epsilon_1)^2\vartheta_1^\prime(0)^2}\frac{\vartheta_1^\prime(\epsilon_1)}{\vartheta_1(\epsilon_1)}\cdot\sum_{a=1}^4\frac{\prod_{f=1}^{2N+8}\vartheta^2_a(m_f)}{\prod_{i}^N\vartheta_a\left(\pm\alpha_i+\frac{\epsilon_1}{2}\right)^2}\cdot\vartheta_a^{\Delta}(x)\notag\\
%
&+\frac{1}{4\eta^{12}\vartheta_1(\epsilon_1)^2\vartheta_1^\prime(0)^2}\frac{\vartheta_2^\prime(\epsilon_1)}{\vartheta_2(\epsilon_1)}
\cdot\frac{\prod_{f=1}^{2N+8}\vartheta_1(m_f)\vartheta_2(m_f)}{\prod_{i=1}^N\vartheta_1\left(\pm\alpha_i+\frac{\epsilon_1}{2}\right)\vartheta_2\left(\pm\alpha_i+\frac{\epsilon_1}{2}\right)}\cdot\left(\vartheta_1^\Delta(x)+\vartheta_2^\Delta(x)\right)\notag\\
&+\frac{1}{4\eta^{12}\vartheta_1(\epsilon_1)^2\vartheta_1^\prime(0)^2}\frac{\vartheta_2^\prime(\epsilon_1)}{\vartheta_2(\epsilon_1)}
\cdot\frac{\prod_{f=1}^{2N+8}\vartheta_3(m_f)\vartheta_4(m_f)}{\prod_{i=1}^N\vartheta_3\left(\pm\alpha_i+\frac{\epsilon_1}{2}\right)\vartheta_4\left(\pm\alpha_i+\frac{\epsilon_1}{2}\right)}\cdot\left(\vartheta_3^\Delta(x)+\vartheta_4^\Delta(x)\right)\notag\\
&+\frac{1}{4\eta^{12}\vartheta_1(\epsilon_1)^2\vartheta_1^\prime(0)^2}\frac{\vartheta_3^\prime(\epsilon_1)}{\vartheta_3(\epsilon_1)}
\cdot\frac{\prod_{f=1}^{2N+8}\vartheta_1(m_f)\vartheta_3(m_f)}{\prod_{i=1}^N\vartheta_1\left(\pm\alpha_i+\frac{\epsilon_1}{2}\right)\vartheta_3\left(\pm\alpha_i+\frac{\epsilon_1}{2}\right)}\cdot\left(\vartheta_1^\Delta(x)+\vartheta_3^\Delta(x)\right)\notag\\
&+\frac{1}{4\eta^{12}\vartheta_1(\epsilon_1)^2\vartheta_1^\prime(0)^2}\frac{\vartheta_3^\prime(\epsilon_1)}{\vartheta_3(\epsilon_1)}
\cdot\frac{\prod_{f=1}^{2N+8}\vartheta_2(m_f)\vartheta_4(m_f)}{\prod_{i=1}^N\vartheta_2\left(\pm\alpha_i+\frac{\epsilon_1}{2}\right)\vartheta_4\left(\pm\alpha_i+\frac{\epsilon_1}{2}\right)}\cdot\left(\vartheta_2^\Delta(x)+\vartheta_4^\Delta(x)\right)\notag\\
&+\frac{1}{4\eta^{12}\vartheta_1(\epsilon_1)^2\vartheta_1^\prime(0)^2}\frac{\vartheta_4^\prime(\epsilon_1)}{\vartheta_4(\epsilon_1)}
\cdot\frac{\prod_{f=1}^{2N+8}\vartheta_1(m_f)\vartheta_4(m_f)}{\prod_{i=1}^N\vartheta_1\left(\pm\alpha_i+\frac{\epsilon_1}{2}\right)\vartheta_4\left(\pm\alpha_i+\frac{\epsilon_1}{2}\right)}\cdot\left(\vartheta_1^\Delta(x)+\vartheta_4^\Delta(x)\right)\notag\\
&+\frac{1}{4\eta^{12}\vartheta_1(\epsilon_1)^2\vartheta_1^\prime(0)^2}\frac{\vartheta_4^\prime(\epsilon_1)}{\vartheta_4(\epsilon_1)}
\cdot\frac{\prod_{f=1}^{2N+8}\vartheta_2(m_f)\vartheta_3(m_f)}{\prod_{i=1}^N\vartheta_2\left(\pm\alpha_i+\frac{\epsilon_1}{2}\right)\vartheta_3\left(\pm\alpha_i+\frac{\epsilon_1}{2}\right)}\cdot\left(\vartheta_2^\Delta(x)+\vartheta_3^\Delta(x)\right)\notag\\
&+\frac{1}{8\eta^{12}\vartheta_1(\epsilon_1)^2\vartheta_1^{\prime}(0)^2}\left(\sum_{a=1}^4\frac{\prod_{f=1}^{2N+8}\vartheta_a(m_f)}{\vartheta_a\left(\pm\alpha_i+\frac{\epsilon_1}{2}\right)}\cdot\vartheta_a^\Delta(x)\right)^2\notag\\
&-\sum_{i}^N\bigg(\frac{1}{2\eta^{12}\vartheta_1(\epsilon_1)\vartheta_1^\prime(0)\vartheta_1(2\alpha_i)^2\vartheta_1(2\alpha_i-2\epsilon_1)\vartheta_1(2\alpha_i-\epsilon_1)^3}\notag\\
&\quad
\cdot\frac{\prod_{f=1}^{2N+8}\vartheta_1\left(\alpha\pm m_f-\frac{\epsilon_1}{2}\right)}{\prod_{j\neq i}\vartheta_1(\alpha_i\pm\alpha_j)\vartheta_1(\alpha_i\pm\alpha_j-\epsilon_1)}
\cdot\left(\vartheta_1^\Delta\left(x-\alpha_i+\frac{\epsilon_1}{2}\right)+\vartheta_1^\Delta\left(x+\alpha_i-\frac{\epsilon_1}{2}\right)\right)\bigg)\notag\\
&\quad + (\alpha_i \rightarrow -\alpha_i)
\notag\\
&+\frac{1}{2\eta^{12}\vartheta_1(2x+2\epsilon_1)\vartheta_1(2x+\epsilon_1)^2\vartheta_1(2x)}\cdot \frac{\prod_{f=1}^{2N+8}\vartheta_1(x\pm m_f+\frac{\epsilon_1}{2})}{\prod_{i=1}^N\vartheta_1(x\pm\alpha_i+\epsilon_1)\vartheta_1(x\pm\alpha_i)}+\left(\epsilon_1\rightarrow -\epsilon_1\right)\,.
\label{eq:2-inst_chi}
\end{align}
}
\end{description}

\section{Wilson loops of 5d $\sprm(N)$ theories}
\label{C}
The Wilson loop expectation value of a 5d gauge theory with ADHM description can be computed by inserting the equivariant Chern character \cite{Shadchin:2004yx, Losev:2003py, Gaiotto:2015una}. For the $\sprm(N)$ case, the $k$-instanton ADHM quantum mechanics is described by two discrete sectors $O(k)_{\pm}$ of the dual group $O(k)$. The partition function of $\sprm(N)_{\theta}+N_f\mathsf{F}$ is
\begin{align}
    Z_{\text{5d}}=1+\sum_{k=1}^{\infty}q_0^kZ_k,\quad\quad\quad\quad\quad Z_k=\begin{cases}
        \frac{1}{2}(Z^{+}_k+Z_k^{-}), &  \theta =0, \\
        \frac{(-1)^k}{2}(Z^{+}_k-Z_k^{-}),  & \theta =\pi,
    \end{cases}
\end{align}
where $q_0$ is the instanton counting parameter, $Z_{k}^{\pm}$ are the $k$-instanton partition functions of the $\pm$ sectors. At the one-instanton level, we have 
\begin{align}
    Z_1^{+}=\frac{\prod_{l=1}^{N_f}\sh(m_l)}{\sh(\epsilon_1)\sh(\epsilon_2)\prod_{i=1}^N\sh(\pm\alpha_i+\epsilon_+)},\quad \quad Z_1^{-}=\frac{\prod_{l=1}^{N_f}\ch(m_l)}{\sh(\epsilon_1)\sh(\epsilon_2)\prod_{i=1}^N\ch(\pm\alpha_i+\epsilon_+)},
\end{align}
where we have used the notation 
\begin{align}
    \sh(x)\equiv e^{x/2}-e^{-x/2},\quad\quad \ch(x)\equiv e^{x/2}+e^{-x/2}.
\end{align}
Higher-instanton expressions can be found in \cite{Kim:2012gu, Hwang:2014uwa}. The partition function with the insertion of the Wilson loop operator can be computed similarly, 
\begin{align}
    W_{\mathbf{R}}=\sum_{k=0}^{\infty}\mathfrak{q}^kW_{k,\mathbf{R}},\quad\quad\quad\quad\quad W_{k,\mathbf{R}}=\begin{cases}
        \frac{1}{2}(W^{+}_{k,\mathbf{R}}+W_{k,\mathbf{R}}^{-}), &  \theta =0, \\
        \frac{(-1)^k}{2}(W^{+}_{k,\mathbf{R}}-W_{k,\mathbf{R}}^{-}),  & \theta =\pi.
    \end{cases}
\end{align}
where $W^{\pm}_{k,\mathbf{R}}$ is computed by inserting the equivariant Chern character. In the fundamental representation, we have
\begin{align}
    \Ch^{\pm}_{k,\text{fund}}(e^\alpha,e^u;q_1,q_2;\chi)=\sum_{i=1}^N(e^{\alpha_i}+e^{-\alpha_i})-(1-q_1)(1-q_2)(q_1q_2)^{-1/2}\sum_{I=1}^n(e^{u_I}+e^{-u_I}\pm\chi),
\end{align}
where $k=2n+\chi$.
Other representations can be generated from the tensor product of the Chern character. For example, in the case of antisymmetric representation $\mathsf{\Lambda}^2$, we have
\begin{align}
    \Ch^{\pm}_{k,\mathsf{\Lambda}^2}(e^\alpha,e^u;q_1,q_2;\chi)=\frac{1}{2}\left[\Ch^{\pm}_{k,\text{fund}}(e^\alpha,e^u;q_1,q_2;\chi)^2-\Ch^{\pm}_{k,\text{fund}}(e^{2\alpha},e^{2u};q_1^2,q_2^2;\chi^2)\right].
\end{align}
Then at zero-instanton level, the Wilson loop expectation value is the character in the representation $\mathbf{R}$, and we have
\begin{align}
    W_{0,\mathbf{R}}=\Ch^{+}_{0,\mathbf{R}}.
\end{align}
At the one-instanton level,
\begin{align}
    W_{1,\mathbf{R}}^{\pm}=\Ch^{\pm}_{1,\mathbf{R}}\cdot Z_1^{\pm}.
\end{align}
The higher-instanton partition functions can be calculated similarly. Lastly, we defined the normalized Wilson loop expectation value
\begin{align}
    \widetilde{W}_{\mathbf{R}}(e^{\alpha},e^{m_l},\mathfrak{q};q_1,q_2)\equiv\frac{{W}_{\mathbf{R}}}{Z_{\text{5d}}}=W_{0,\mathbf{R}}+(W_{1,\mathbf{R}}-Z_{1}\cdot W_{0,\mathbf{R}})q_0+\mathcal{O}(q_0^2).
\end{align}

Note that the ADHM quantum mechanic calculations generically only compute the partition function and Wilson loop expectation values for $\sprm(N)+N_f\mathsf{F}$ for $N_f\leq 2N+4$ \footnote{However, the one-instanton calculation is correct for $N_f= 2N+5$ and $2N+6$.}, higher number of fundamental flavors can be calculated using the method of the blowup equations \cite{Kim:2021gyj}.
\clearpage

\bibliography{reference} 
\bibliographystyle{JHEP}

\end{document}